\documentclass[preprint]{aastex}

\usepackage{graphicx}
\usepackage{natbib}
\usepackage{epsfig}
\usepackage[usenames,dvipsnames]{xcolor}
\usepackage[normalem]{ulem}
\usepackage{cancel}

\begin{document}

\shorttitle{An improved model for collisional growth in protoplanetary disks}
\shortauthors{Garaud et al.}


\title{From dust to planetesimals: an improved model for collisional growth in protoplanetary disks}

\author{P. Garaud$^{1}$, F. Meru$^{2,3}$ M. Galvagni$^{4}$, C. Olczak$^{5,6,7}$} 
\affil{$^1$ Department of Applied Mathematics and Statistics, UC Santa Cruz, 1156 High Street Santa Cruz, \\
$^2$ Institut f\"ur Astronomie, ETH Z\"urich, Wolfgang-Pauli-Strasse 27, 8093 Z\"urich, Switzerland\\
$^3$ Institut f\"ur Astronomie und Astrophysik, Universit\"at T\"ubingen, Auf der Morgenstelle 10, 72076 T\"ubingen, Germany\\
$^4$ Institute of Theoretical Physics, University of Zurich, Winterthurerstrasse 190, 8057 Zurich, Switzerland, \\
$^5$ Astronomisches Rechen-Institut (ARI), Zentrum f{\"u}r Astronomie Universit{\"a}t Heidelberg, M{\"o}nchhofstrasse 12-14, 69120 Heidelberg, Germany, \\
$^6$ Max-Planck-Institut f{\"u}r Astronomie (MPIA), K{\"o}nigstuhl 17, 69117 Heidelberg, Germany, \\
$^7$ National Astronomical Observatories of China, Chinese Academy of Sciences (NAOC/CAS), 20A Datun Lu, Chaoyang District, Beijing 100012, China}


\begin{abstract}
   Planet formation occurs within the gas- and dust-rich environments of protoplanetary disks. Observations of these objects show that the growth of
    primordial submicron-sized particles into larger aggregates occurs at the earliest evolutionary stages of the disks. However, theoretical models of particle growth 
    that use the Smoluchowski equation to describe collisional coagulation and fragmentation have so far failed to produce large particles while maintaining a significant 
    population of small grains. This has been generally attributed to the existence of two barriers impeding growth due to bouncing and fragmentation of colliding particles. 
    In this paper, we demonstrate that the importance of these barriers has been artificially inflated through the use of simplified models 
    that do not take into account the stochastic nature of the particle motions within the gas disk. 
    We present a new approach in which the relative velocities
    between two particles is described by a probability distribution function that models both deterministic motion (from the vertical settling, radial drift and azimuthal drift) 
   and stochastic motion (from Brownian motion and turbulence).  Taking both into account can give quite different results to what has been considered recently in other studies.
    We demonstrate the vital effect of two ``ingredients'' for particle growth: the proper implementation of a velocity distribution
    function that overcomes the bouncing barrier and, in combination with mass transfer in high-mass-ratio collisions, boosts the growth of larger
    particles beyond the fragmentation barrier. A robust result of our simulations is the emergence of two particle populations (small and large), 
    potentially explaining simultaneously a number of long-standing problems in protoplanetary disks, including planetesimal formation close to the central star, the presence of mm- to cm-size
    particles far out in the disk, and the persistence of micron-size grains for millions of years.
\end{abstract}



\section{Introduction}
\label{sec:intro}

\subsection{Observations confront theory}

Protoplanetary disks are universally thought to be the birthplace of planets. Furthermore, the ubiquity of terrestrial to gas-giant exoplanets
discovered around other stars (\citet{Cassan2012,Howard2012}, see also the compilation at http://www.circumstellardisks.org) strongly
suggests that the planet formation process must be robust and efficient.  This simple fact, however, has been until now rather difficult to reconcile
with a number of observational properties of protoplanetary disks.
 
Consider for instance a scenario in which planets are created by the gradual coagulation of small dust grains into progressively larger particles, eventually leading
to the formation of planetesimals and then planets. This idea turns out to be surprisingly hard to model within the constraints set by disk
observations.  Indeed, signatures of ongoing gas accretion onto the host star typically disappear after 10\,Myr \citep{Calvet2000}. In the core-accretion scenario \citep{Pollack1996},  coagulation must be fast enough to drive growth from
submicron-sized dust grains to earth-sized rocky planets while gas is still present in sufficient quantities to form gas giants.

 More crucially, particles as they grow must necessarily pass through a critical phase where their radial velocity (induced by gas drag) approaches a small but significant fraction of their orbital speed \citep{Weidenschilling1977}. They are prone to drift into the central star within 10 to a 100 orbits unless growth through that phase is rapid enough. In short, coagulation must be {\it very} efficient for planets to exist. 

However, protoplanetary disks are usually detected via a characteristic near- and mid-infrared emission that is in excess of the expected pure photospheric flux of their host stars \citep{CohenKuhi1979}. Objects showing such a feature in their spectral energy distribution are known as low-mass T-Tauri or intermediate-mass Herbig AeBe stars \citep{GillettStein1971,Strom1972}. The main contribution to the excess emission comes from hot micron-sized dust grains located within a few AU of the star. The coagulation
efficiency of these tiny dust grains, upon collision, is usually thought to be much higher than that of larger bodies \citep{DullemondDominik2005}. This implies that the growth process would quickly render them invisible at these wavelengths, shifting the excess emission to longer wavelengths, in contrast to what is observed \citep{2001AJ....121.2065H,2007ApJ...667..308C,2009ApJ...705..314S}.

A very similar problem was raised with the observational discovery of ``large'' grains at large orbital radii away from the central star -- more precisely, mm-to-cm-size grains, at radii at least as far out as a few tens of AU \citep{1998Natur.392..788H,Wilner2005,Rodmann2006,Lommen2009}. Forming such grains at these orbital distances, where the dynamical timescale is more than a hundred times longer than around 1AU, is just as difficult as forming meter-sized objects in the inner stellar system. Furthermore, mm- and cm-size grains at 30AU are also subject to rapid inward drift. Again, coagulation has to be very efficient for growth to cm-size to happen despite these two problems, but cannot a priori be too efficient otherwise the outer (and inner) disk would disappear entirely on a much shorter timescale than the one observed \citep{Garaud2007,2008ApJ...686.1195H}.

A possible way out of this impasse was proposed by \citet{DullemondDominik2005}, who introduced the possibility of fragmentation in addition to
coagulation. Fragmentation is a common outcome of high-velocity collisions \citep[see review by][]{BlumWurm2008}. As long as the fragmentation
rate is high enough, and the size distribution of the fragments is sufficiently steep to favor the creation of small grains upon impact, then it is
possible to maintain a healthy micron-size grain population at all times in the disk \citep{DullemondDominik2005,Brauer2008}. Unfortunately,
while fragmentation solves one problem it introduces another -- particle growth beyond a new ``fragmentation barrier'', corresponding to cm-size at
a few AU and down to less than mm-size at 30AU, seems to be impossible unless additional physics are taken into account \citep{Brauer2008,Johansen2008}.

\subsection{Modeling particle growth using ensemble models}

Upon realization of the true complexity of this problem, much effort was dedicated to modeling the collisional dynamics of particles in
protoplanetary disks in more detail. The ensemble approach traditionally used to study aerosol growth in planetary atmospheres
\citep{KovetzOlund1969,Podolak1980} and grain growth in molecular clouds
\citep{Rossi1991,Ossenkopf1993}, in which one models the evolution of the particle size distribution function with the Smoluchowski
coagulation/fragmentation equations \citep{Smoluchowski1916,Melzak1957}, has become one of the most commonly used techniques applied to the
problem. We shall adopt it here too.

The Smoluchowski equations are integro-differential equations. Information on the collisional dynamics of particles must be provided in the form of ``collisional kernels'', which summarize all aspects of the collisions, including the relative velocities of the colliding particles, their cross sections, their sticking and/or fragmentation probabilities, the fragment distribution, etc... Formally speaking, the coagulation kernel of two particles ``$i$'' and ``$j$'' (the latter being two indices that reference the particle masses $m_i$ and $m_j$  for instance) should be written as:
\begin{eqnarray} 
K_{ij} && = \int_{{\cal S}_i} d{\bf S}_i  \int_{{\cal S}_j} d {\bf S}_j \int_{\cal I} d{\bf I} \int_0^\infty d \Delta_{ij} \int_0^1 d \epsilon_{ij}^s \left\{  \left[ \Delta_{ij} \epsilon^s_{ij} \right]  \right. \nonumber \\
&& \left. p_\epsilon(\epsilon^s_{ij}| {\bf S}_i,{\bf S}_j,\Delta_{ij}, \delta_{ij};{\bf D}) p_\Delta(\Delta_{ij}|{\bf S}_i,{\bf S}_j;{\bf D}) p_I({\bf I}|{\bf S}_i,{\bf S}_j;{\bf D})   p_{S_i}({\bf S}_i;{\bf D}) p_{S_j}({\bf S}_j;{\bf D})  \right\}
\label{eq:crazy}
\end{eqnarray}
or, in other words, a high-dimension multivariate integral over 
\begin{itemize}
\item all possible variations in each particle's internal properties that result in the same particle mass, loosely summarized here as the vector
  ${\bf S}$, which may include shape, composition, porosity, or other, and is drawn from a multivariate probability distribution function
  (p.d.f. hereafter) $p_S$. Note that if the masses of the two particles are different then the p.d.f.s for each of the particles may also be
  different. Since the grain composition could depend on the local temperature, the p.d.f.s should also depend on the local disk properties, loosely
  described here as the vector ${\bf D}$. Finally, $p_S$ may also be a function of time, if the repeated collisions lead to internal changes in the
  particles -- this has not be explicitly written here for simplicity.
\item all possible impact configurations, loosely summarized here as the vector ${\bf I}$, which includes impact parameter and solid angle, and drawn from a p.d.f. $p_I$. If collisional outcomes are independent of ${\bf I}$, then the integral separates and yields the mean area cross section of the collision $\bar a_{ij}$. 
\item all possible relative velocities $\Delta_{ij}$ (drawn from a p.d.f. $p_\Delta$) for the collisions of the two particles. Note that since particle motion is induced by collisional/frictional interaction with the gas, this p.d.f. depend on the respective structure of both particles, and on the disk model considered. 
\item all possible sticking efficiencies $\epsilon^s_{ij}$ for the collisions of the two particles, drawn from a p.d.f. $p_\epsilon$. This p.d.f depends on everything else, including the local disk properties. 
\end{itemize}
The integrand in the square brackets is the
product of the collisional relative velocity $\Delta_{ij}$, and the sticking efficiency $\epsilon^s_{ij} $, discussed above.

Clearly, equation~(\ref{eq:crazy}) is {\it far} too complicated to be of practical use. \citet{Tanaka2005} and \citet{DullemondDominik2005}
dramatically simplified the kernel expression to a product of three means, which may themselves only depend on mean quantities (and on the disk
model):
\begin{equation}
K_{ij} = \bar a_{ij}(\bar {\bf S}_i,\bar {\bf S}_j;{\bf D}) \bar \Delta_{ij}(\bar {\bf S}_i,\bar {\bf S}_j;{\bf D}) \bar \epsilon^s_{ij} (\bar {\bf S}_i,\bar {\bf S}_j,\bar \Delta_{ij};{\bf D}) \mbox{   , } 
\label{eq:Kernelold}
\end{equation} 
where $\bar a_{ij}$ is the mean collisional cross section of a pair of particles of mass $m_i$ and $m_j$, assuming mean internal properties $\bar {\bf S}_i$ and $\bar {\bf S}_j$; 
$\bar \Delta_{ij} $ is a mean relative velocity and $\bar \epsilon^s_{ij}$ is the mean sticking efficiency of the collisions. A similar equation is used to approximate the fragmentation kernel.

\subsection{Improving collisional dynamics models}

Having generally adopted this much simpler formulation, studies of the evolution of the particle size distribution function then focused on refining parametric prescriptions for the mean quantities involved. Progress was made on all three fronts -- in characterizing the collisional velocities, the particle structure (which determines the collisional cross-section) and collisional outcomes. 

Thanks to the vast advances in supercomputing, it is now possible to study the dynamics of a vast number of inertial particles interacting with the
disk turbulence. Using this information, one can then
infer statistical properties of individual particle velocities \citep{NelsonGressel2010,Carballido2011}, and of pairwise relative velocities
\citep{Carballido2008,Carballidoal2010}. The latter have helped begin to validate a theoretical prescription for the rms relative particle velocity of
colliding particles proposed by \citet{OrmelCuzzi2007}, which is commonly used in dust coagulation models to construct the quantity $\bar \Delta_{ij}$
used in equation~(\ref{eq:Kernelold}). However, the study of the p.d.f.s of collisional velocities is still very much in its infancy (see
\citet{Carballidoal2010} for a first attempt at the problem using direct simulations of turbulence, and \citet{Hubbard2012} using shell models for turbulence).

In parallel, much work has also been done to improve our understanding of the outcome of particle collisions, and typical particle structures (the
latter often depending on the collisional history) using both experimental and numerical methods. It is generally accepted that collisions involving
only micron-size particles have very high sticking efficiency and that successive collisions result in the formation of fractal aggregates (see
\citealt{Ossenkopf1993}; also see the review by
\citealt{BlumWurm2008}). The collisions of these aggregates with one another (or with monomers) then lead to further sticking (with very high
probability), but also to compaction \citep{DominikTielens1997,BlumWurm2000}. The resulting objects are roughly mm- or sub-mm size grains that have
varying degrees of porosity \citep{Blum2006}. In short, for any collision involving only sub-mm particles, one can usually adopt a mean sticking
efficiency $\bar \epsilon_{ij}^s \simeq 1$; the calculation of the area cross-section as a function of the particle mass, by contrast, is more
complicated if one wishes to take into account all aforementioned effects \citep{Ossenkopf1993,Ormel_porosity,Suyama_porosity_model,Okuzumi_porosity_model,2010A&A...513A..57Z}.

Collisions involving larger particles have much more varied outcomes, depending on their internal properties, relative velocities, and mass
ratio. Given the sheer dimensionality of parameter space, and the 35 orders of magnitude mass-range between mm-size grains and Earth-size objects, a
comprehensive study of the problem is impossible. It is generally agreed that sticking continues to occur for low-velocity (alternatively low-energy)
collisions, that intermediate velocity (energy) collisions may lead to bouncing with compaction \citep{BlumMunch1993,Weidling2009}, and that the
outcome of high-velocity collisions very much depends on the mass ratio of the two particles -- with equal mass collisions being much more likely to
lead to complete fragmentation than unequal mass collisions \citep[see the review by][]{BlumWurm2008}. In fact, high-velocity collisions in high-mass
ratio events can also lead to partial or complete sticking \citep{Teiser_Wurm_highVcoll,Kothe2010}. 
If the fraction of the projectile mass that sticks to the target is larger than the total
amount of mass ejected away on impact -- a phenomenon called ``mass transfer'' hereafter --, then high-mass-ratio collisions can lead to the growth of
the target. 

However, the details of the transitions between
each of these regimes is highly dependent on the particle masses and their internal structure, on the impact parameter and on the collisional energy.
\citet{Guttler2010} recently proposed a complex model for sticking, bouncing and fragmentation
  outcomes  that summarizes a suite of laboratory experiments.  Numerical simulations of larger, centimeter-sized, dust aggregates show that the fragmentation
  velocities may be larger than the previously assumed 100\,cm/s \citep{Four_population} and that the outcome may be dependent on the inhomogeneity of
  the aggregate \citep{Geretshauser_inhomogeneity}, the target and projectile sizes as well as the porosity of the aggregates (Meru et al, in
  prep).

\subsection{Proposed work}

A number of papers have been published since 2005 which gradually take into account more and more of the aforementioned effects, as well as additional
disk physics \citep{Birnstiel2010}, whilst keeping the kernel formulation given in equation~(\ref{eq:Kernelold}).

Without giving an exhaustive list, we note in particular the work of \citet{Brauer2008} and \citet{Windmark2012b}. \citet{Brauer2008} incorporate 
the collisional dust dynamics with a whole-disk model to study simultaneously the evolution of the particle size distribution function through 
coagulation and fragmentation as well as mean motions. They model the mean sticking and fragmentation probabilities using piecewise linear or piecewise parabolic functions, 
and consider sticking and fragmentation, but no bouncing. They also include effects such as ``cratering'' (or equivalently, erosion\footnote{Note that \citet{Guttler2010} refer to this mechanism as ``erosion'' while \citet{Windmark2012b} use these terms interchangeably.}), in which high-mass-ratio collisions only lead to the partial rather than the complete
  fragmentation of the larger body. Later on, \mbox{\citet{Windmark2012b}} discussed even more
  complex models, which include sticking, fragmentation, cratering {\it and} bouncing, taking into account the mass dependence of the regime boundaries \citep{Weidling_expts}. Furthermore, they also considered the possibility of ``mass transfer'',  where high-mass-ratio collisions lead to the net growth of the target. \citet{Windmark2012b} showed that the latter could lead to the growth of particles well-beyond the fragmentation barrier, but only if they are introduced -- somewhat artificially -- beyond it to start with. 
Nevertheless, despite all these added physics, most models
fail to robustly produce planetesimals at 1AU and cm-size particles far out in the disk, for reasonable disk assumptions, whilst keeping a micron-size
dust grain population consistent with observations.

By contrast, the manner in which the coagulation and fragmentation kernels are constructed has not really been revisited since 2005. However, the statement that they can be approximated using products of three separately-taken averages is only technically correct if the quantities considered are mutually independent -- which is certainly not the case. We demonstrate in this paper that the failure of previous models to reconcile disk observations with exoplanetary observations simply stems from the use of equation~(\ref{eq:Kernelold}), which vastly oversimplifies the problem and results in misleading conclusions concerning  particle growth in protoplanetary disks. 

Our goal here is not to propose a {\it complete} new model for the construction of collisional kernels 
-- certainly, none as complicated as the one suggested in equation~(\ref{eq:crazy}). Instead, we aim to demonstrate that adding even just one element of stochasticity to the model, taking into account the full p.d.f. of relative velocities instead of their mean only, resolves a number of the aforementioned problems, at least qualitatively speaking. 

To our knowledge, \citet{Okuzumi2011} were the first to take into account velocity p.d.f.s explicitly in collisional dust growth models 
with application to protoplanetary disks. They focused on studying very small particles only, which interact via Brownian motion 
and mean drift. Here, we also consider larger particles, and include the effect of turbulence in driving stochastic motions.  We loosely follow here the same general methodology first introduced by \citet{Okuzumi2011} and later in the ISIMA (International Summer
Institute for Modeling in Astrophysics) report by\citet{Galvagni2011} and  \citet{Windmark2012}. 
We construct a plausible p.d.f. for the relative velocities of two particles interacting with the gaseous disk. We then use the
latter to construct new collisional kernels, and study the resulting collisional particle growth and fragmentation. Our work nevertheless differs
notably from previously published ideas, in several ways. First, we explicitly take into account the effects of turbulence in constructing the velocity p.d.f, by
contrast with \citet{Okuzumi2011}. Secondly, while \citet{Galvagni2011} and \citet{Windmark2012} did take turbulence into account, they modeled the relative velocities
p.d.f. as a Maxwellian, which neglects the difference between chaotically-driven particle motion (e.g. Brownian motion and turbulence) 
and regular particle motion (radial drift, azimuthal drift and vertical settling induced by gas drag).  Moreover, the model
proposed here does take drift into account more carefully, and is much more generally applicable.  Finally, we point out and correct a mathematical
error in the work of \citet{Windmark2012}, which affects their coagulation and fragmentation kernels.

In what follows, we present in Section~\ref{sec:general} the general framework associated with the use and solution of the Smoluchowski equations. The
construction of the kernels is discussed in Section~\ref{sec:kernels}, and contrasts the traditional approach with our new proposed
one. Section~\ref{sec:diskmodel} presents a sample disk model, and Section~\ref{sec:results} shows how the evolution of the particle size distribution
function in this disk dramatically changes between the two approaches. The principal finding is that particle growth beyond the traditional bouncing and
fragmentation barriers is possible, and is a robust feature of all models that include relative velocity p.d.f.s in the kernel construction {\it and}
consider the possibility of mass transfer in high-mass-ratio collisions. The particle size distribution function rapidly evolves into two
populations, one composed of small dust grains and one composed of much larger particles. The quantitative details of these two
populations, however, are strongly model-dependent. This is discussed in Section~\ref{sec:discuss}, together with recommendations for future work
directions.

\section{General framework}
\label{sec:general}

We consider a local region of a protoplanetary disk, 
centered around the mid-plane at orbital radius $r$. The properties of the gas disk are assumed to be known, steady, and independent of the dust properties. This is a good approximation as long as the dust-to-gas mass ratio is not too large (which we will assume here), and as long as the evolution timescale of the dust 
size distribution function is short compared with the disk evolution timescale (which needs to be verified a posteriori for the disk model selected). 

The particle size (or mass) distribution varies with time, in a manner that is controlled both by the flux of particles in and out of the region, and
through collisional coagulation or fragmentation. In this paper we neglect the former, for simplicity (the same approximation was used by
\citealt{Windmark2012}). We discuss the implications of this assumption in Section~\ref{sec:discuss}. In this local model, the evolution of the
particle size distribution function is therefore simply governed by the Smoluchowski coagulation-fragmentation equations. We now present these equations for
completeness, but essentially follow the work of \citet{Brauer2008}.

\subsection{Coagulation-Fragmentation equations}
\label{sec:smolueqs}

Assuming that the masses of the dust particles can take any values among a {\it discrete} ``mass-mesh'' 
$\{m_1,m_2, ... m_I\}$, we define $N_i$ as the number density of particles of mass $m_i$. The coupled
nonlinear evolution equations for the functions $N_i(t)$ are the discrete form of the Smoluchowski coagulation-fragmentation equations
\citep{Smoluchowski1916,Melzak1957}, in which, for $i=1..I$,
\begin{eqnarray}
\frac{d N_i}{dt} = \frac{1}{2} \sum_{j,k=1}^I C_{jki} K_{jk} N_j N_k  -
\sum_{j=1}^I K_{ij} N_i N_j  \nonumber \\+
\frac{1}{2}\sum_{j,k=1}^I F_{jk} N_j N_k  N_{ijk}^f -
\sum_{j=1}^I F_{ij} N_i N_j \mbox{   . } 
\label{eq:smolufrag}
\end{eqnarray}
In each of these equations, the first two terms on the right-hand-side model coagulation, while the last two terms model fragmentation. The quantities
$K_{ij}$ and $F_{ij}$ are the coagulation and fragmentation kernels respectively, which contain crucial information
on the relative velocities of particles of mass $m_i$ and $m_j$, their collisional cross-section,
and their probability of sticking and fragmenting as discussed in Section \ref{sec:intro}. 
The construction of these kernels is discussed in Sections~\ref{sec:ingredients} and \ref{sec:kernels}. 

The first term on the right-hand-side of equation~(\ref{eq:smolufrag}) models the increase in the number density of particles of mass $m_i$ resulting from the
coagulation of particles of mass $m_j$ and $m_k$ such that $m_j + m_k \simeq m_i $.  If the mass mesh is linearly spaced, then there exists a
mass-point $m_i$ which is {\it exactly} equal to $m_j + m_k$, namely the point for which $i=j+k$. In that case, this coagulation term reduces to the
more commonly-used $ \frac{1}{2} \sum_{j=1}^{i} K_{j,i-j} N_j N_{i-j}$.

In order to follow the growth of particles across many orders of magnitude in mass, however, one cannot realistically use a linearly-spaced mass-mesh -- it is much more common to use a logarithmically-spaced one instead. With a non-uniform mesh, however, 
$m_j + m_k$ usually falls in between two existing consecutive mass-points $m_{i-} $ and $m_{i+} $, with $m_{i-} < m_j + m_k < m_{i+}$. This leads to the introduction of the third-rank tensor $C_{jki}$, which assigns different probabilities for the coagulation of $m_j$ and $m_k$ into either $m_{i-} $ or $m_{i+} $. The details of 
how to construct $C_{jki}$ in such a way as to conserve mass and coagulation rates were first discussed by \citet{KovetzOlund1969}, and are very nicely summarized in \cite{Brauer2008}. Following their work, we take
\begin{eqnarray}
&& C_{jki} = p_{jk} \mbox{ if } i = i- \nonumber \\
&& C_{jki} = (1-p_{jk}) \mbox{ if } i = i+ \nonumber \\
&& C_{jki} = 0 \mbox{ otherwise} \nonumber 
\end{eqnarray}
where 
\begin{equation}
p_{jk} = \frac{m_{i+}  - (m_j + m_k) }{m_{i+}- m_{i-}} \mbox{   . } 
\end{equation}

The second and fourth term on the right-hand-side of equation~(\ref{eq:smolufrag}) describe how particles of mass $m_i$ disappear when they collide with particles of mass $m_j$, and either coagulate (second term) or fragment (fourth term).

Finally, the third term on the right-hand-side of equation~(\ref{eq:smolufrag}) describes how $N^f_{ijk}$ particles of mass $m_i$ are created when particles $m_j$ and $m_k$ (with $m_j+m_k > m_i$) collide and fragment. 
In what follows, we assume that in any fragmentation event the number density of fragments generated is
a power law with fixed index $-\xi $, where $\xi = 11/6 = 1.83$ as in \cite{Brauer2008} for instance. Since the mass bins are logarithmically distributed, we have
\begin{equation}
N^f_{ijk} = A_{jk}  m_i^{1-\xi} H(L-i) \mbox{   , } 
\end{equation}
where $H(L-i)$ is a discrete Heaviside function (defined so that
$H(L-i) = 0$ if $i>L$), $L$ is the index of the mass-point $m_L$, which is the mass of
the largest fragment generated, and the normalization $A_{jk}$ is uniquely defined 
by conservation of mass during the collision:
\begin{equation}
m_j + m_k = A_{jk}  \sum_{i=1}^{L}  m_i^{2-\xi} \mbox{   . } 
\end{equation}
We assume here for simplicity that $m_L$ is either equal to the largest mass-point still
satisfying $m_L < m_j+m_k$, or, in the occasional very high-mass collisions
events with $m_j+m_k \ge m_I$ (where $m_I$ is the largest mass-point
in our mesh), $m_L$ is chosen to be equal to $m_I$. 

It can be shown that this set of equations conserves the total mass of particles in
the system exactly, with 
\begin{equation}
\frac{d}{dt} \sum_{i=1}^I m_i N_i =0 \mbox{   . } 
\label{eq:masscons}
\end{equation}
Numerical truncation errors in the evaluation of the sums, however, can cause the total mass to drift (see below).

The coagulation-fragmentation equations are intrinsically nonlinear, and -- aside from very specific kernels and initial conditions -- must be evolved forward in time numerically.  This system 
of equations is also inherently very stiff, so it is crucial to use an implicit time-stepping
method. Here, we have chosen to use a simple Euler first-order time-centered
scheme, and calculate the Jacobian of the right-hand-side
explicitly. We use an automatic adaptive time-stepping scheme to
ensure the required level of precision. The algorithm was successfully tested against well-known analytical solutions of these equations, reviewed by \citet{Aldous1999}. 

We continuously check for conservation of the total mass. As in \citet{Brauer2008}, when the ratio of the smallest to the largest particle masses considered falls below the numerical precision of the system (which is the case in some of the simulations presented below), mass conservation becomes difficult to enforce. When the total mass drifts by more than an acceptable margin (typically, one part in a million) away from the initial mass, the difference is added back to the mass distribution function in the form of monomers at the smallest size considered. Comparison of the results using this simple fix, with those obtained using a quad-precision real arithmetic (which yields reliable results up to much larger particle sizes) shows that the difference is minimal. Since quad-precision real arithmetic requires about 10 times longer integration times, we use the fix in all simulations shown below.

Finally, let us first discuss a subtle but rather important point concerning the discrete version of the Smoluchowski equations. In the derivations presented so far, we have defined $N_i$ to be the number density of particles of mass $m_i$. This definition is useful for numerical purposes, but can be troublesome when interpreting the results. Indeed, consider for instance a uniformly distributed mass-mesh with a total of 100 mass-points, and a number density of $N_0$ particles per cm$^3$. If the particle mass distribution function is itself uniform, then $N_i = N_0/100$ for all $i$. By contrast, had we selected 200 mass-points instead, i.e. a finer mass-mesh, then $N_i = N_0/200$ for all $i$. This simple example illustrates that $N_i$ is a resolution-dependent quantity. 

A much better approach is to consider the continuous particle mass distribution function $dn/dm$, defined 
so that, for any two masses $m_1$ and $m_2$, $\int_{m_1}^{m_2} (dn/dm) dm$ is the number density of particles of mass between $m_1$ and $m_2$. This definition is now entirely independent of the mass-discretization used. In order to relate $dn/dm$ to the numerically determined $N_i$'s, simply note that, as long as there are many mass-points between $m_1$ and $m_2$, $\int_{m_1}^{m_2} (dn/dm) dm \simeq \sum_{i = i_1}^{i_2} N_i $ where $i_1$ is the index of the closest mass-point to $m_1$, and similarly for $i_2$. Looking at the integral as a Riemann sum, one can then straightforwardly identify
\begin{equation}
N_i = \left. \frac{dn}{dm}\right|_{m=m_i}  dm_i = \left. \frac{dn}{dm}\right|_{m=m_i}  (m_i - m_{i-1}) \mbox{   . } 
\label{eq:dndm}
\end{equation}
In Section~\ref{sec:results}, we present all our results in terms of $dn/dm$; the latter are calculated from the discretized $N_i$ using equation~(\ref{eq:dndm}).

\subsection{Ingredients for coagulation and fragmentation}
\label{sec:ingredients}

While the solution of the Schmoluchowski coagulation-fragmentation equation
poses interesting mathematical and computational problems, most of the physical complexity 
of the system studied manifests itself in the construction of the coagulation and fragmentation 
kernels, $K_{ij}$ and $F_{ij}$. The latter must be constructed for each $(i,j)$ 
pair (i.e. for all collisions involving one particle of mass $m_i$ and one particle of mass $m_j$). As discussed in Section~\ref{sec:intro}, they have traditionally been simplified into 
\begin{equation}
K_{ij} = \bar a_{ij} \bar \Delta_{ij} \bar \epsilon^s_{ij} \mbox{   and } F_{ij} = \bar a_{ij} \bar \Delta_{ij} \bar \epsilon^f_{ij}  \mbox{   , } 
\label{eq:KernelB}
\end{equation} 
where $\bar a_{ij}$ is the mean collisional cross-section of $(i,j)$ particle pairs (averaged over all possible individual internal structures resulting in the same mass), $\bar \Delta_{ij}$ is an estimate of the mean relative velocity of all $(i,j)$ pairs (averaged over all possible realizations), and $ \bar \epsilon^s_{ij}$ is the mean
sticking probability of a collisional event, given this mean structure and at that mean relative velocity (and similarly for $\bar \epsilon^f_{ij}$). 
This formulation is equivalent to assuming that {\it all} $(i,j)$ pairs 
have the {\it same} collisional cross-section (neglecting the possibility of variations in density through compaction, and variations in shape), collide with the {\it same} relative velocity $\bar \Delta_{ij}$ (ignoring the chaotic nature of particle motions), and that all these collisions have the {\it same} sticking and fragmentation probabilities $ \bar \epsilon_{ij}^s$ and $\bar \epsilon_{ij}^f $ (neglecting variations with impact parameter and inherent variability in collisional outcomes).  

Often used with specific oversimplified prescriptions for the fragmentation and sticking probabilities $\bar \epsilon^s_{ij} ( \bar \Delta_{ij} )$ and
$\bar \epsilon^f_{ij} ( \bar \Delta_{ij} )$, this formulation has, over the years, led to the introduction and discussion of arguably artificial
``bouncing'' and ``fragmentation'' barriers. As we shall demonstrate, however, taking into account even a single stochastic component -- namely the
chaotic nature of the particle motion -- in the description of the collisions overcomes these barriers, and can explain the
growth of large particles whilst retaining a population of small particles quite naturally.  In what follows, we therefore still adopt a simple
collisional cross-section model {\it and} a simple model for the fragmentation and sticking probabilities in {\it individual} collisions, but look at
the effects of modeling the stochastic nature of the particle's relative velocities. We now describe these ingredients in more detail. 

\paragraph{Collisional cross-section.}

The collisional cross-section of two particles $i$ and $j$ 
is typically taken to be their area cross-section, although 
electrostatic or gravitational effects could be important for tiny dust particles and planetesimals respectively. The calculation of an
area cross-section for two given particles requires knowledge of their shape and internal density, which both depend on their
collisional history. Taking these effects into account properly can only be done via Monte Carlo simulations, as in \citet{ZsomDullemond2008} for instance. In ensemble evolution models (such as the one presented here) one is forced instead to make simplifying assumptions. Following most previous work on the subject \citep{Tanaka2005,DullemondDominik2005}
we will consider only spherical particles of solid density $\rho_s$, and use
\begin{equation}
a_{ij} = \bar a_{ij} = \pi (s_i+ s_j)^2 \mbox{   , } 
\label{eq:aij}
\end{equation}
and $s_i$ is the assumed radius of a particle of mass $m_i$. The two are related via $m = \frac{4}{3} \pi \rho_s s^3$, where $\rho_s = 1$g/cm$^3$ as in \citet{Windmark2012}. 

\paragraph{Sticking, Bouncing and Fragmenting.}

The complex dependence of possible collisional outcomes on the collisional velocity and particle properties was discussed in Section~\ref{sec:intro}, and is still very much the subject of ongoing investigations. However, since our purpose is not to give quantitatively accurate predictions for the evolution of the particle size distribution function, but rather to study qualitatively the impact of the introduction of our new model, we use the simplest possible prescription for the fragmentation and sticking probabilities. As in \citet{Windmark2012} we assume that, in any specific collisional event, particles stick for $\Delta_{ij} < v_b$ and fragment if $\Delta_{ij} > v_f$. Between $v_b$ and $v_f$ lies the ``bouncing'' region. Hence we select, for individual collisions, 
\begin{eqnarray} 
\epsilon_{ij}^f = H(\Delta_{ij} - v_f) \mbox{   , } \nonumber \\
\epsilon_{ij}^s = H(v_b - \Delta_{ij} ) \mbox{   , } 
\label{eq:epsbasic}
\end{eqnarray}
where $H$ is a Heaviside function. In all that follows, we use for the sake of comparison with the work of \citet{Windmark2012}, the values $v_b = 5$cm/s (unless otherwise specified) and $v_f  = 100$cm/s. 

\paragraph{Relative velocity.}

The various possible sources of relative motion for solid particles in protoplanetary disks, as reviewed by
\citet{WeidenschillingCuzzi1993} for instance, can be divided into two categories: regular motion caused by frictional interaction with the mean
component of the gas velocity (usually divided into radial drift, azimuthal drift, and vertical settling), and chaotic motion caused by collisions
with gas molecules (Brownian motion) and interaction with the turbulent component of the gas velocity.

Let us begin by considering the motion of particles in a gas at rest (i.e. supposing that the macroscopic gas velocity is zero in the frame of reference considered). 
Each particle undergoes Brownian motion via collisions with the gas molecules, and thereby acquires a random velocity, which is isotropic and has an amplitude drawn from a Maxwellian p.d.f.. Since the velocities of two particles both undergoing Brownian motion are uncorrelated, the p.d.f. of their relative velocity $\Delta_{ij}$ is also a Maxwellian with
\begin{equation}
p_{\rm B}(\Delta_{ij}) =  \sqrt{\frac{2}{\pi}} \frac{\Delta_{ij}^2}{(\sigma^{\rm B}_{ij})^3} \exp \left( - \frac{\Delta_{ij}^2}{2(\sigma^{\rm B}_{ij})^2 }\right) \mbox{   , } 
\end{equation}
where $(\sigma^{\rm B}_{ij})^2 = kT(m_i + m_j)/m_i m_j$, $k$ is the Boltzmann constant and $T$ is the local gas temperature. The mean value of the relative collisional velocities is 
\begin{equation}
\bar \Delta_{ij}^B  = \sqrt{ \frac{8 kT (m_i + m_j)}{\pi m_i m_j }} = \sqrt{\frac{8}{\pi}} \sigma^{\rm B}_{ij}\mbox{   . } 
\label{eq:meanMaxwell}
\end{equation}
Note that, as defined, $\sigma^{\rm B}_{ij}$ is exactly $1/\sqrt{3}$ times the rms collisional velocity and $\bar \Delta_{ij}^B$ is $\sqrt{8/3\pi}$ times the rms collisional velocity. This implies that for a Maxwellian, to a good approximation, the mean and rms velocities are equal to one another -- a well-known result that we shall use later.

In protoplanetary disks, however, particles can have significant mean velocities with respect to the gas. This is particularly true for the larger particles, which 
orbit around the star at near-Keplerian speeds, while the azimuthal gas velocity is notably sub-Keplerian (since the gas is pressure-supported while the particles are not). As a result of this differential motion, collisions with gas molecules transfer net momentum to the particles. This effect is often modeled as an added drag term in the particle's equation of motion, and leads to radial and azimuthal drift within the disk \citep{Whipple1972,Weidenschilling1977,Nakagawa1986}, as well as net vertical settling toward the mid-plane over time \citep{GoldreichWard1973,Dubrulle1995,Garaudal2004}.  

Let's consider a frame that is rotating with the local Keplerian angular velocity $\Omega_{\rm K}$. In that frame, 
assuming that the surface density of solids is much smaller than the surface density of the gas, particles
of mass $m_i$ have mean radial and azimuthal
velocities $\bar u_i$ and $\bar v_i$ respectively, with 
\begin{eqnarray}
&& \bar u_i = \frac{1}{S^2_i + 1} \left( u_g -  2 S_i\eta v_{\rm K} \right) \mbox{   , }  \nonumber \\
&& \bar v_i = \frac{1}{2S_i } ( \bar u_i - u_g ) \mbox{   , } 
\end{eqnarray}
where $u_g$ is the radial gas velocity, and $\eta$ is related to the deviation of the gas azimuthal velocity from the local azimuthal Keplerian velocity $v_{\rm K} = r \Omega_{\rm K}$. All these quantities depend on the selected disk model, and are presented in more detail in Section~\ref{sec:diskmodel}. Finally, $S_i$ is the Stokes number of particles of mass $m_i$, defined as
\begin{equation}
S_i = \tau_i \Omega_{\rm K} = \frac{s_i \rho_s}{\rho_m c}  \Omega_{\rm K} = \frac{ \sqrt{2\pi} s_i \rho_s }{\Sigma }  \mbox{   , } 
\end{equation}
where $\tau_i$ is the particle stopping time, $\rho_m$ is the mid-plane disk density, $c$ is the local sound speed and $\Sigma$ is the local surface density of the gas (see Section~\ref{sec:diskmodel}). Note that we have assumed here that particles remain in the Epstein gas drag regime, which is true for small particles but may not be accurate above a certain size. This assumption should be dropped should one require a quantitatively more accurate prediction for the evolution of the particle size distribution function, but is satisfactory within our stated qualitative goals (see Section~\ref{sec:intro}). 

The mean settling velocity of a particle of mass $m_i$ is calculated as in \citet{Birnstiel2010}:
\begin{equation}
\bar w_i   =  - h_i \Omega_{\rm K} \min\left( S_i, \frac{1}{2} \right) \mbox{   , } 
\end{equation}
where $h_i$ is the vertical scaleheight of particles of mass $m_i$, and where the minimum function 
guarantees that the vertical velocity is at most equal to that of the epicyclic motion. 
The particle disk scaleheight $h_i$ can be obtained by 
solving an advection-diffusion equation \citep{Dubrulle1995,Garaud2007}, which yields
\begin{equation}
\frac{1}{h_i^2}    = \frac{ S_i \Omega_K }{D_i} + \frac{1}{h^2}\mbox{   , } 
\end{equation}
where $h$ is the local gas disk scaleheight, $D_i = \frac{\nu}{1+S_i}$ is the reduced turbulent diffusivity for the particles of size $s_i$, and $\nu$ is the local turbulent viscosity of the gas (see Section~\ref{sec:diskmodel}). 

As should be obvious from these derivations, particles of different sizes have different velocities relative to that of the gas. As a result, they also acquire relative velocities with respect to each other. This time, however, no random effects are involved so the mean value of the relative velocities of particles $i$ and $j$, both induced by gas drag, is
\begin{equation}
\bar \Delta^D_{ij} = \sqrt{ (\bar u_i - \bar u_j)^2 + (\bar v_i - \bar v_j)^2 + (\bar w_i - \bar w_j)^2}\mbox{   , } 
\label{eq:meanvel}
\end{equation}
and the p.d.f. is well approximated by a $\delta-$function centered around the mean. 

However, the effect of gas drag described above is only strictly valid in a ``non-turbulent accretion disk'' -- somewhat of an oxymoron. As first
described by \citet{Volkal1980}, particles interacting with turbulent eddies have inherently stochastic motions. Their velocity p.d.f. is
non-isotropic (when the turbulence itself is anisotropic), and depends sensitively on the particle stopping time compared with the eddy turnover time
at the energy dissipation and injection scales \citep{Volkal1980,OrmelCuzzi2007}. To add to the complexity of the problem,
particle trajectories are no longer necessarily independent of one another -- two small particles trapped in the same eddy have
strongly correlated motion. Recent progress has been made to characterize the problem using full
numerical simulations \citep{Carballidoal2010} and simplified vortex gas models \citep{RastPinton2011}, although the results still
have limited applicability in both cases. A theoretical model of the rms
relative velocities in turbulence, which we assume is very similar to their mean relative velocity $\bar \Delta_{ij}^T$ by analogy with the case of
Brownian motion, was proposed by \citet{OrmelCuzzi2007} (see their equations~(27)-(29)): 
\begin{eqnarray}
\bar \Delta_{ij}^{T} &=&  \sqrt{\frac{S_i - S_j}{S_i + S_j}}\left[ \frac{S_i^2}{S_i + {\rm Re}^{-1/2}} - \frac{S_j^2}{S_j + {\rm Re}^{-1/2}}   \right]^{1/2} v_e \mbox{			if			} S_{i,j} \le {\rm Re}^{-1/2} \mbox{   , }  \nonumber \\
\bar \Delta_{ij}^{T}&=& \left[ 2.2 S_i  - S_j + \frac{2 S^2_i}{S_i + S_j} \left( \frac{1}{2.6} + \frac{S_j^3}{1.6 S_i^3 + S_i^2 S_j} \right) \right] ^{1/2}  v_e \mbox{ if  }  {\rm Re}^{-1/2}  \le  S_i \le  1 \mbox{ for all } j<i  \mbox{   , } \nonumber \\
\bar \Delta_{ij}^T &=& \left[ \frac{1}{1 + S_i} + \frac{1}{1 + S_j}  \right]^{1/2} v_e   \mbox{   , } \mbox{     		if		      } 1 < S_i  \mbox{ for all } j<i 
\label{eq:meanturbvel}
\end{eqnarray}
where Re is the local Reynolds number, which we will take to be $10^8$ (see Section \ref{sec:diskmodel}), $v_e = \sqrt{\alpha} c$ is the typical eddy velocity, and $\alpha$ is the
turbulent intensity parameter (see Section \ref{sec:diskmodel}). Note that in these equations we have assumed that $S_i \ge S_j$. If the opposite is
true, then the indices should be switched.

\section{Kernel prescriptions -- old vs. new}
\label{sec:kernels}

All the information described in the previous Section now needs to be combined to create the coagulation and fragmentation kernels. In this section, 
we contrast the ``old approach'', which includes all ensemble models of particle growth until 2011, and the ``new approach'' first proposed by \citet{Okuzumi2011}, \citet{Galvagni2011} and \citet{Windmark2012}, which we build upon.

\subsection{Traditional approach} 
\label{sec:oldkernels}

As discussed in the previous section, the collisional and fragmentation kernels are usually given by equation~(\ref{eq:KernelB}). The problem then reduces
to calculating a single value for the mean relative velocities of the two particles $ \bar \Delta_{ij}$, combining information from all the possible sources of 
motion described in Section~\ref{sec:ingredients}. Traditionally, $\bar \Delta_{ij}$ is constructed as in \citet{Tanaka2005}, with 
\begin{equation}
\bar \Delta_{ij} = \sqrt{ (\bar \Delta^B_{ij})^2 +  (\bar \Delta^D_{ij})^2 + (\bar \Delta^T_{ij})^2 } \mbox{   , } 
\label{eq:tradbardelta}
\end{equation}
and the fragmentation and coagulation probabilities $\bar \epsilon^s_{ij}$ and $\bar \epsilon^f_{ij}$ are simply taken to be 
\begin{equation}
\bar \epsilon_{ij}^s = \epsilon_{ij}^s ( \bar \Delta_{ij}) \mbox{   and   } \bar \epsilon_{ij}^f = \epsilon_{ij}^f ( \bar \Delta_{ij}) \mbox{   , } 
\label{eq:tradeps}
\end{equation}
where the functions $\epsilon_{ij}^s$ and $\epsilon_{ij}^f$ are constructed (for instance\footnote{In most prior work to date, more elaborate prescriptions for $\epsilon_{ij}^s$ and $\epsilon_{ij}^f$ are used, which include piecewise linear or piecewise parabolic functions.}) as in equation~(\ref{eq:epsbasic}).

\subsection{New approach} 
\label{sec:newkernels}

In order to take into account the stochastic nature of the particle velocities, collisional and fragmentation kernels need to be re-written\footnote{Note that here we have ignored the variability in the area cross-section, as discussed earlier.} as
\begin{eqnarray}
&& K_{ij} = \bar a_{ij} \int_0^\infty \Delta_{ij} p(\Delta_{ij}) \epsilon^s_{ij}(\Delta_{ij}) d \Delta_{ij} \mbox{   , } \nonumber \\
&& F_{ij} = \bar a_{ij} \int_0^\infty \Delta_{ij} p(\Delta_{ij}) \epsilon^f_{ij}(\Delta_{ij}) d\Delta_{ij}  \mbox{   , } 
\label{eq:newkernels}
\end{eqnarray} 
which one may recognize as a much-simplified version of equation~(\ref{eq:crazy}), where $p(\Delta_{ij})$ is a single p.d.f. for the relative velocities of the two particles (dropping the suffix $\Delta$ on the $p$ for clarity of notation), which ought to combine information from all the possible sources of motion described in Section~\ref{sec:ingredients}. The main difficulty here comes from the fact that, while the p.d.f.s of particles undergoing Brownian motion or mean drift are known, we still do not have any knowledge of the shape of the p.d.f. of relative velocities for particles undergoing collisions via turbulent motions. One is left to chose the latter somewhat arbitrarily.

\citet{Galvagni2011} and \citet{Windmark2012} 
both proposed that $\Delta_{ij}$ should be distributed as a Maxwellian. 
\citet{Galvagni2011} considered all possible regimes simultaneously, and constructed the Maxwellian so that its mean is given by $\bar \Delta_{ij}$ as defined in equation~(\ref{eq:tradbardelta}). \citet{Windmark2012} only considered the turbulent regime, and required that the mean of the Maxwellian be $\bar \Delta^T_{ij}$ instead. Both models naturally yield the same p.d.f. in a regime that is dominated by turbulent motions (i.e. when $\bar \Delta^T_{ij} \simeq \bar \Delta_{ij} \gg \bar \Delta^D_{ij} $) but both models fail in the limit where the system is dominated by regular motion. The proposed model by \citet{Galvagni2011} overestimates the particle dispersion in that limit, while \citet{Windmark2012} do not address the question. 

In what follows, we build on the work of \citet{Okuzumi2011} and propose an alternative, more rigorous approach for the construction of $p(\Delta_{ij})$, which takes into account the fact that individual particles have deterministic velocities (induced by radial drift,
azimuthal drift and vertical settling), {\it in addition to} stochastic motions induced by Brownian motion and turbulence. 
For simplicity, in all that follows we assume that the stochastic motions are isotropic\footnote{This assumption is not
  absolutely necessary, but is very useful. Without it, many of the integrals involved in the derivation of the p.d.f of the relative velocities have to
  be evaluated numerically.}.

Let's focus first on a given direction (taking here, for the sake of illustration, the radial direction).
We model the one-dimensional velocity p.d.f. of particle $i$ as a Gaussian with mean $\bar u_i$ (the mean radial drift velocity discussed earlier) and standard deviation $\sigma_i$ 
(specified later). Hence,
\begin{equation}
p_r(u_i) = \frac{1}{\sqrt{2\pi} \sigma_i } \exp\left(-\frac{(u_i-\bar u_i)^2}{2\sigma_i^2}\right) \mbox{   , } 
\end{equation}
where the index $r$ denotes the radial direction. Note that, in this context, $u_i$ can be positive or negative. 
Given a second particle indexed by $j$, with a similarly defined radial velocity p.d.f., a well-known (although non-trivial) result from probability theory is that the p.d.f. of their relative radial velocities is 
\begin{equation}
p_r (u_i - u_j) = p_r(\Delta_{r,ij}) = \frac{1}{\sqrt{2 \pi} \sigma_{ij}} \exp \left[ - \frac{\left(\Delta_{r,ij}  -(\bar{u}_{i} - \bar{u}_{j})\right)^2}{2 \sigma_{ij}^2} \right]\mbox{   , } 
\end{equation}
where $\sigma_{ij}^2 = \sigma_i^2 + \sigma_j^2$.
Note that $\Delta_{r,ij} = u_i - u_j$ can be positive or negative. Furthermore, it changes sign upon permutation of $i$ and $j$. 

Similar expressions can be obtained for the p.d.f.s of relative velocities in the azimuthal ($\varphi$) and vertical ($z$) directions. Assuming independence of the distributions in each respective direction, the p.d.f. of the three-dimensional relative velocity is simply the product of the three derived p.d.f.s: 
\begin{equation}
p_{3D}({\mathbf \Delta}_{ij} ) d {\mathbf \Delta }_{ij}  = p_r(\Delta_{r,ij})  p_\varphi(\Delta_{\varphi,ij}) p_z(\Delta_{z,ij})   d \Delta_{r,ij} d \Delta_{\varphi ,ij}  d \Delta_{z,ij}   \mbox{   . } 
\label{eq:p3d}
\end{equation}
From this expression, with a little bit of algebra (see Appendix A) and assuming isotropy in the particle stochastic motion (i.e. assuming that $\sigma_i$ and $\sigma_j$ are independent of the direction selected), one can show that the p.d.f. of the {\it amplitude} of the relative velocity $\Delta_{ij}$ = $|{\mathbf \Delta}_{ij}|$ is 
\begin{equation}
p(\Delta_{ij} )  = \frac{1}{\sqrt{2\pi} \sigma_{ij}} \frac{\Delta_{ij}}{\bar \Delta^D_{ij}} \left[ \exp\left( -\frac{(\Delta_{ij} - \bar \Delta^D_{ij})^2 }{ 2\sigma_{ij}^2} \right) - \exp\left(-\frac{(\Delta_{ij} + \bar \Delta^D_{ij})^2 }{ 2\sigma_{ij}^2} \right) \right] \mbox{   , } 
\label{eq:newpdf}
\end{equation}
where $\bar \Delta^D_{ij} $ is given by equation~(\ref{eq:meanvel}). This time, both $\Delta_{ij}$ and $\bar \Delta^D_{ij}$ are always positive by construction, and invariant under permutation of $i$ and $j$. It can be shown that this p.d.f. is always positive, and that the integral over all possible values of $\Delta_{ij} $ is indeed 1, as required. 

In the limit where $\bar \Delta^D_{ij} \ll \sigma_{ij}$, which corresponds to cases where the deterministic relative velocities of the particles are much smaller than their rms velocities (i.e. in the case where relative motion is dominated by Brownian motion or turbulent motion), $p(\Delta_{ij})$ reduces to a Maxwellian distribution (to lowest order in a Taylor expansion in $\bar \Delta^D_{ij}$),  with 
\begin{equation}
p(\Delta_{ij} ) \simeq \sqrt{\frac{2}{\pi}} \frac{\Delta_{ij}^2}{\sigma_{ij}^3} \exp \left( - \frac{\Delta_{ij}^2}{2\sigma_{ij}^2} \right) \mbox{   , } 
\label{eq:newpdf2}
\end{equation}
which has a mean 
\begin{equation}
\bar \Delta_{ij} = \sqrt{\frac{8}{\pi}} \sigma_{ij}\mbox{   . } 
\label{eq:Maxwellmean}
\end{equation}
In order to ensure that the mean velocity of that distribution is the same as that obtained from Brownian motion or turbulent motion when these dominate the dynamics of the system, we take
\begin{equation}
\sigma^2_{ij} = \frac{\pi}{8} \left[ \left( \bar \Delta^B_{ij} \right)^2 + \left( \bar \Delta^T_{ij} \right)^2 \right] \mbox{   , } 
\label{eq:sigmaij}
\end{equation}
where $\bar \Delta^B_{ij}$ and  $\bar \Delta^T_{ij}$  are defined in equations~(\ref{eq:meanMaxwell}) and (\ref{eq:meanturbvel}) respectively. Our formalism recovers the idea proposed by \citet{Galvagni2011} and by \citet{Windmark2012} in the limit where turbulence is the dominant factor. Note that, as defined, $\sigma_{ij}$ is not the rms velocity of the distribution defined in equation~(\ref{eq:newpdf}), although it is related to its variance. In the limit where the mean drift velocity is close to zero, $\sigma_{ij}$ is, as discussed earlier, $1/\sqrt{3}$ times the rms velocity. 
Our derivation is so far very similar to that of \citet{Okuzumi2011}, and recovers their relative velocity p.d.f. in the limit where the particle dispersion is dominated by Brownian motion.

When $\bar \Delta^D_{ij} \gg \sigma_{ij}$, which correspond to cases where the deterministic relative velocities of the particles are much larger than their rms velocities, the second Gaussian term in equation~(\ref{eq:newpdf}) is negligible, and the relative velocity p.d.f. reduces to 
\begin{equation}
p(\Delta_{ij} ) \simeq \frac{1}{\sqrt{2\pi} \sigma_{ij}} \frac{\Delta_{ij}}{\bar \Delta^D_{ij}} \exp\left( -\frac{(\Delta_{ij} - \bar \Delta^D_{ij})^2 }{ 2\sigma_{ij}^2} \right) \mbox{   . } 
\label{eq:newpdf1}
\end{equation}
It it interesting to note that this is not a Gaussian, but has a somewhat more significant tail for larger values of $\Delta_{ij}$. 

\begin{figure}[h]
\centerline{\includegraphics[width=0.5\textwidth]{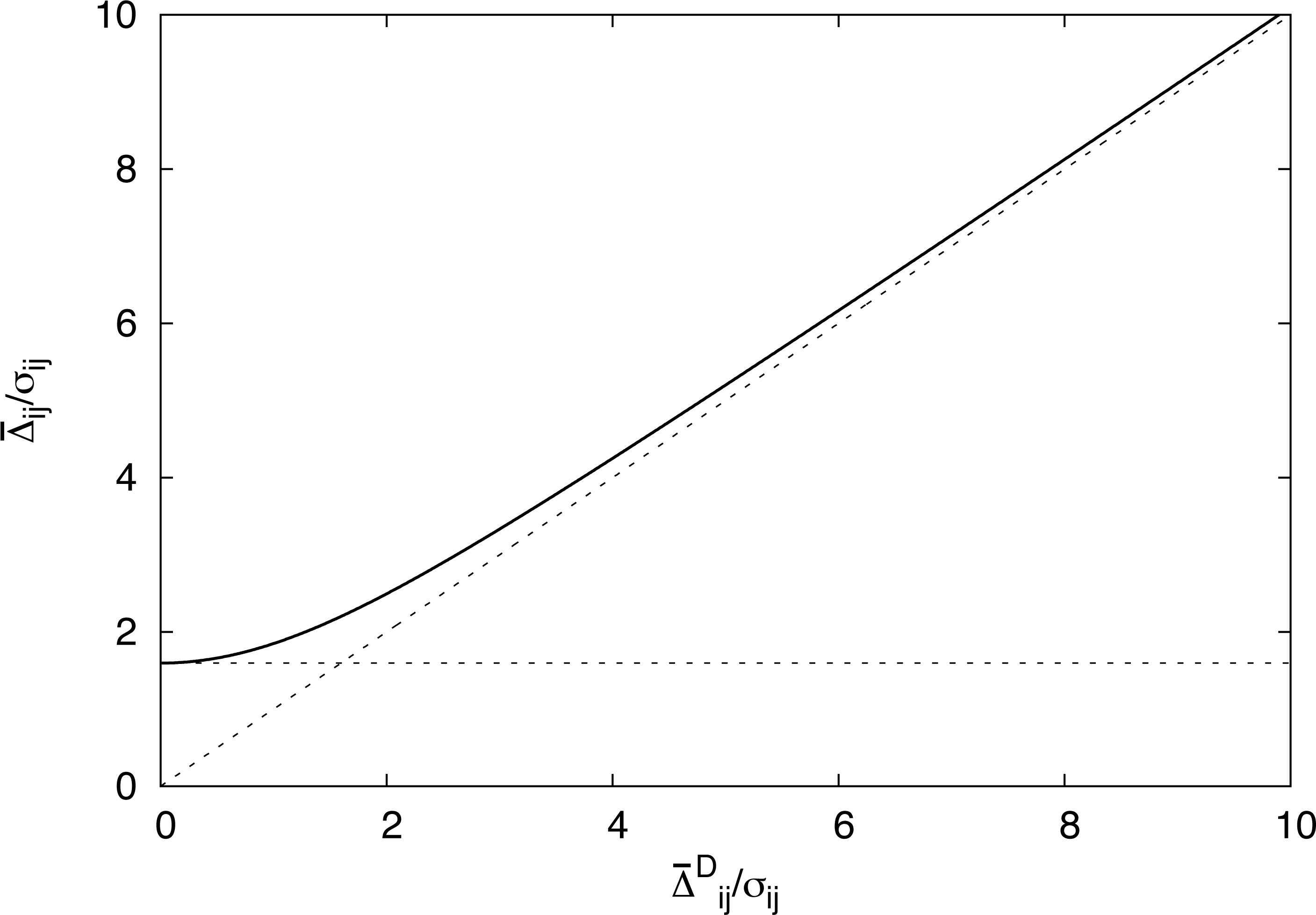}}
\caption{Comparison between the mean particle velocity and the mean drift velocity. The solid line shows $\bar \Delta_{ij}/\sigma_{ij}$. The slanted dotted line is the $y=x$ line, showing that $\bar \Delta_{ij} \rightarrow \bar \Delta^D_{ij}$ when $\Delta^D_{ij}\gg \sigma_{ij}$. The horizontal line is at $\bar \Delta_{ij} = \sqrt{8/\pi} \sigma_{ij}$.}
\label{fig:Edij}
\end{figure}

It is worth noting that the mean relative drift velocity, $\bar \Delta^D_{ij}$, is not equal to the actual mean relative velocity $\bar \Delta_{ij}$; the latter is given by 
\begin{equation}
\bar \Delta_{ij} = \int_0^\infty \Delta_{ij} p(\Delta_{ij})  d \Delta_{ij} = \frac{ \sigma_{ij}^2 + (\bar \Delta^D_{ij})^2}{\bar \Delta^D_{ij}} {\rm erf}\left( \frac{\bar \Delta^D_{ij}}{\sqrt{2} \sigma_{ij}}\right) +\sqrt{\frac{2}{\pi} } \sigma_{ij} \exp\left( - \frac{(\bar \Delta^D_{ij})^2}{2 \sigma_{ij}^2} \right) \mbox{   . } 
\label{eq:dijevalue}
\end{equation}
The relationship between $\bar \Delta^D_{ij}$ and $\bar \Delta_{ij}$ is shown in Figure~\ref{fig:Edij}. We see that $\bar \Delta_{ij}$ tends to $\bar \Delta^D_{ij}$ when $\bar \Delta^D_{ij} \gg \sigma_{ij}$, i.e. 
when the mean particle velocity is large compared with the rms particle velocity, and  to $\sqrt{8/\pi} \sigma_{ij}$ when $\bar \Delta^D_{ij} \ll \sigma_{ij}$, i.e. when the mean particle velocity is small compared with the rms particle velocity. This behavior is consistent with expectations. 

Finally, we can now calculate the new coagulation and fragmentation kernels, following equation~(\ref{eq:newkernels}), by convolving the velocity p.d.f. constructed in equation~(\ref{eq:newpdf}) with the individual collision sticking and fragmentation probabilities. With the simple expressions for $\epsilon_{ij}^s$ and $\epsilon_{ij}^f$ given by equation~(\ref{eq:epsbasic}), it is possible to calculate these kernels analytically\footnote{For other more complex prescriptions for $\epsilon_{ij}^{s,f}$, the integrals need to be performed numerically.}
\begin{eqnarray}
&& F_{ij} = \bar a_{ij} \int_0^\infty \Delta_{ij} p(\Delta_{ij}) \epsilon_{ij}^f(\Delta_{ij}) d\Delta_{ij} = \bar a_{ij} \int_{v_f}^\infty \Delta_{ij} p(\Delta_{ij}) d\Delta_{ij}  \nonumber \\
&& = \bar a_{ij} \left\{ \frac{(\bar \Delta^D_{ij})^2 + \sigma_{ij}^2}{2 \bar \Delta^D_{ij}} A(v_f) +  \frac{\sigma_{ij}}{\sqrt{2\pi} \bar \Delta^D_{ij}} B(v_f)  \right\} \mbox{   , } 
\label{eq:fij}
\end{eqnarray}
and
\begin{eqnarray}
&& K_{ij}  = \bar a_{ij} \int_0^{\infty} \Delta_{ij} p(\Delta_{ij}) \epsilon_{ij}^s(\Delta_{ij})   d \Delta_{ij}  = \bar a_{ij} \int_0^{v_b} \Delta_{ij} p(\Delta_{ij}) d \Delta_{ij} 
\label{eq:kij}  \\
&& = \bar a_{ij} \left\{ \frac{(\bar \Delta^D_{ij})^2 + \sigma_{ij}^2}{2 \bar \Delta^D_{ij}} \left[ 2 {\rm erf} \left( \frac{ \bar \Delta^D_{ij} }{\sqrt{2} \sigma_{ij}} \right) - A(v_b) \right] +  \frac{\sigma_{ij}}{\sqrt{2\pi} \bar \Delta^D_{ij}} \left[ 2 \bar \Delta^D_{ij} \exp \left( - \frac{ (\bar \Delta^D_{ij})^2}{2\sigma_{ij}^2} \right) - B(v_b) \right]  \right\}\mbox{   , }  \nonumber 
\end{eqnarray}
where for simplicity of notation we have defined 
\begin{eqnarray}
&& A(v) = {\rm erf} \left( \frac{v + \bar \Delta^D_{ij}}{\sqrt{2} \sigma_{ij}} \right) - {\rm erf} \left( \frac{v - \bar \Delta^D_{ij}}{\sqrt{2} \sigma_{ij}} \right)  \mbox{   , } \nonumber \\
&& B(v) =(v + \bar \Delta^D_{ij}) \exp \left( - \frac{  (v- \bar \Delta^D_{ij})^2}{2\sigma_{ij}^2}\right) - (v- \bar \Delta^D_{ij}) \exp \left( - \frac{  (v + \bar \Delta^D_{ij})^2}{2\sigma_{ij}^2}\right)  \mbox{   . } 
\end{eqnarray}
Note that if $\bar \Delta^D_{ij} \ll \sigma_{ij}$ then (to lowest order in a Taylor expansion in $\bar \Delta^D_{ij}$)
 \begin{eqnarray} 
F_{ij} = \bar a_{ij} \bar \Delta_{ij} \frac{2\sigma^2_{ij} + v_f^2}{2\sigma^2_{ij}}  \exp \left( - \frac{ v_f^2}{2\sigma_{ij}^2}\right) \mbox{   , } \nonumber \\
K_{ij} =  \bar a_{ij} \bar \Delta_{ij} \left[  1 -\frac{2\sigma^2_{ij} + v_b^2}{2\sigma^2_{ij}}  \exp \left( - \frac{ v_b^2}{2\sigma_{ij}^2}\right) \right]  \mbox{   . } 
\label{eq:Maxwell}
\end{eqnarray}

If desired, one can thereby construct ``mean'' sticking and fragmentation probabilities
\begin{eqnarray}
&& \bar \epsilon_{ij}^s = \frac{K_{ij}}{\bar a_{ij}  \bar \Delta_{ij}} = \frac{  \int_0^\infty \Delta_{ij} p(\Delta_{ij}) \epsilon_{ij}^s(\Delta_{ij})   d \Delta_{ij} }{ \int_0^\infty \Delta_{ij} p(\Delta_{ij}) d \Delta_{ij} } \mbox{   , } \nonumber \\
&& \bar \epsilon_{ij}^f = \frac{F_{ij}}{\bar a_{ij}  \bar \Delta_{ij}}  =  \frac{  \int_0^\infty \Delta_{ij} p(\Delta_{ij}) \epsilon_{ij}^f(\Delta_{ij})   d \Delta_{ij} }{ \int_0^\infty \Delta_{ij} p(\Delta_{ij}) d \Delta_{ij} }\mbox{   , } 
\label{eq:meaneps}
\end{eqnarray}
so that, using these expressions, we have an exact analog for equation~(\ref{eq:KernelB}). Note that the second term in each of these expressions is a rather general formula for the mean sticking and fragmentation probabilities, which is valid for any assumed velocity p.d.f., and any assumed individual collision and fragmentation probabilities, but does assume that the collisional outcomes are independent of the particle density and shape. 

It can easily be verified that when $\epsilon_{ij}^s(\Delta_{ij}) + \epsilon_{ij}^f(\Delta_{ij}) =1$ for individual particles, then we also have $\bar \epsilon_{ij}^s + \bar \epsilon_{ij}^f  = 1$, or equivalently, 
\begin{equation}
K_{ij}  + F_{ij}  = \bar a_{ij} \bar \Delta_{ij}\mbox{   . } 
\end{equation}
This happens for instance when $v_b = v_f$ (i.e. in the absence of bouncing) in the simple Heaviside model given by equation~(\ref{eq:epsbasic}), and
can be verified directly by adding equations~(\ref{eq:fij}) and (\ref{eq:kij}).

Note that the coagulation kernel derived in equation (\ref{eq:kij}) looks like the one presented by \citet{Okuzumi2011} in their equation (12), but is slightly different because the 
sticking probability of individual collisions in their paper is taken to be a piecewise linear function of the collisional energy rather than a Heaviside function of the collisional velocity. 
Our results would of course recover one another had we chosen the same sticking probability function, and in the limit where the particle dispersion is not affected by turbulence (i.e. in the limit where $\bar \Delta^T_{ij} = 0$). By contrast, \citet{Windmark2012} would not recover the same formula in any limit.

Indeed, note that \citet{Windmark2012} discuss the role of the \emph{total} sticking and fragmentation probabilities (see their equations (4) and (5)), in our notation written
  as
\begin{equation}
  [\epsilon_{ij}^s] = \int_0^{v_b} p_M(\Delta_{ij}) d\Delta_{ij} \quad \mbox{and} \quad [\epsilon_{ij}^f] = \int_{v_f}^\infty p_M(\Delta_{ij})
    d\Delta_{ij} \, ,
\end{equation}
(where $p_M$ denotes a Maxwellian distribution) rather than the \emph{mean}  sticking and fragmentation velocities $\bar \epsilon_{ij}^s$ and $\bar \epsilon_{ij}^f$. However, as should be clear from the derivation presented above, 
these ``total'' probabilities play no role in the calculation of the kernels -- the means must be used instead. Their results should therefore be considered incorrect even though they look qualitatively similar to ours (see Section \ref{sec:results}). 

\begin{figure}
\centerline{\includegraphics[width=\textwidth]{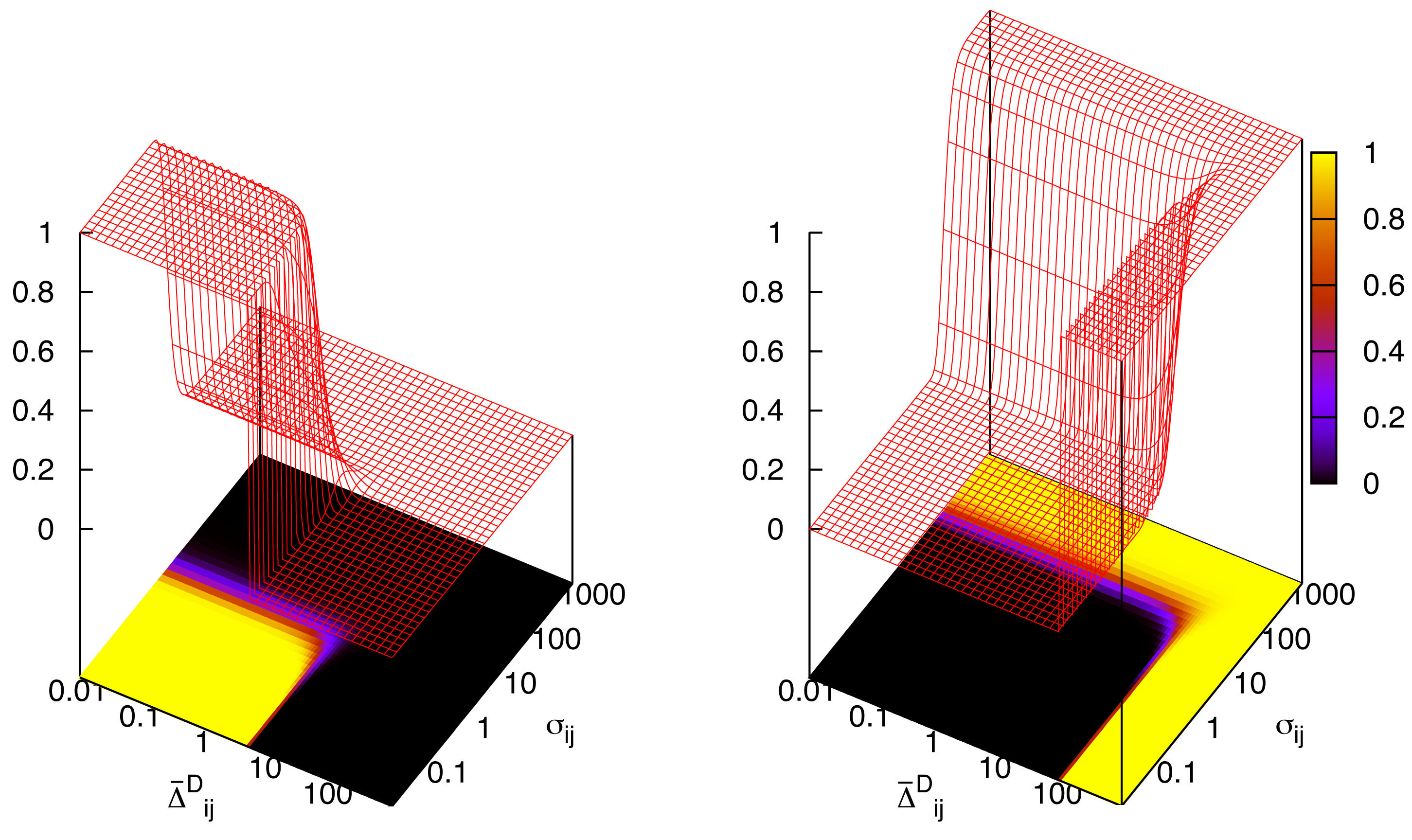}}
\caption{Effective sticking and fragmentation probabilities, assuming a bouncing threshold of $v_b = 5$cm/s and a fragmentation threshold of $v_f =
  100$cm/s, and calculated using equations~(\ref{eq:meaneps}). The left-hand plot shows shows $\bar \epsilon_{ij}^s$ and the right-hand plot shows
  $\bar \epsilon_{ij}^f$. The color bar is the same for both plots. Note how perfect sticking requires that both $\sigma_{ij}$ and $\bar
  \Delta^D_{ij}$ should be smaller than $v_b$. Also note how the transition from perfect sticking to perfect fragmentation, across the ``bouncing
  region'', is much smoother when increasing $\sigma_{ij}$ at fixed $\bar \Delta^D_{ij}$ than increasing $\bar \Delta^D_{ij}$ at fixed
  $\sigma_{ij}$. The Maxwellian-only approach (cf. \citealt{Galvagni2011} and \citealt{Windmark2012}) assumes that $\bar \Delta^D_{ij} = 0$.}
\label{fig:epsij}
\end{figure}

Figure~\ref{fig:epsij} shows $\bar \epsilon_{ij}^s$ and $\bar \epsilon_{ij}^f$, as calculated in equation~(\ref{eq:meaneps}), as a function of $\sigma_{ij}$
and $\bar \Delta_{ij}^D$, for $v_b = 5$cm/s and $v_f = 100$cm/s (see Section~\ref{sec:ingredients}). We see that when both $\sigma_{ij}$ {\it and}
$\bar \Delta_{ij}^D$ are small compared with the bouncing threshold, sticking is very probable, while when {\it either} $\sigma_{ij}$ {\it or} $\bar
\Delta_{ij}^D$ are large compared with $v_f$, fragmentation is very probable. The transition between the two regimes, however, is much smoother when
$\sigma_{ij}$ is large than when it is small; this behavior is, again, consistent with expectations.

It is important to note that the effective sticking and fragmentation probabilities $\bar \epsilon_{ij}^s$ and $\bar \epsilon_{ij}^f$ are never zero, by contrast with the individual collision functions $\epsilon_{ij}^s$ and $\epsilon_{ij}^f$. This result is very general, and stems from the fact that $\bar \epsilon_{ij}^s$ and $\bar \epsilon_{ij}^f$ are convolution integrals of $\epsilon_{ij}^s$ and $\epsilon_{ij}^f$ with the assumed particle velocity p.d.f.s times the particle relative velocity. 
As a result, fragmentation is always possible (albeit unlikely) even when the mean relative velocity of two particles is very small, and sticking is always possible (albeit unlikely) even when the mean relative velocity of two particles is very large.  This has a number of important consequences, as shown in Section~\ref{sec:results}. 

\section{Disk model and consequences for particle dynamics}
\label{sec:diskmodel}

Since the mean and stochastic relative velocities of the particles depend on the disk model and the
location considered we present them here for completeness, and discuss them in the light of the previous section.

\subsection{Disk model}

We assume that the surface density of the gas follows a  truncated power law: 
\begin{equation}
\Sigma(r) = \frac{M}{2\pi R r} e^{-r/R}\mbox{   , } 
\end{equation}
where $M$ is the disk mass, $r$ is the orbital radius, and $R$ is the truncation radius.  As discussed by \citet{LyndenBellPringle1974} (see also
\citealt{Hartmann1998}; \citealt{Garaud2007}), this model has the desirable property of being an attracting self-similar solution of the equations
describing the ``viscous spreading'' of the disk, as long as the turbulent viscosity $\nu$ varies linearly with radius:
\begin{equation}
 \nu(r) = \nu_{\rm  AU} \frac{r}{1{\rm AU}} \mbox{   , } 
\label{eq:nur}
\end{equation}
where the subscript AU refers to a quantity measured at 1 AU. 
Conservation of angular momentum implies that the radial
velocity of the gas is 
\begin{equation}
u_g(r) = -\frac{3}{r^{1/2} \Sigma(r)} \frac{\partial}{\partial r} \left(
  r^{1/2} \nu(r) \Sigma(r) \right) = - 3 \frac{\nu_{\rm AU}  }{1{\rm AU}}  \left( \frac{1}{2} - \frac{r}{R} \right) \mbox{   , } 
\label{eq:ug}
\end{equation}
while the radial force balance near the mid-plane implies that the azimuthal velocity of the gas is
\begin{equation}
v_g(r) = v_{\rm K}(r) + \frac{1}{2\rho_m(r) \Omega_K(r) } \frac{\partial p_m}{\partial r} = (1-\eta(r)) v_{\rm K}(r) \mbox{   , } 
\label{eq:vg}
\end{equation}
where $p_m$ is the mid-plane gas pressure, and where 
\begin{equation}
\eta(r) = - \frac{1}{2r \rho_m(r) \Omega^2_{\rm K}(r)} \frac{\partial p_m}{\partial r} \mbox{   . } 
\end{equation}
As illustrated by equations~(\ref{eq:ug}) and (\ref{eq:vg}), 
the local gas velocity depends on the local thermodynamical structure of the disk. We now turn to describing the latter in more detail. 

Assuming that the disk is thin, vertically isothermal and in hydrostatic
equilibrium implies that the density profile is
\begin{equation}
\rho(r,z) = \rho_m(r) \exp \left( - \frac{z^2}{2 h^2(r)} \right) \mbox{   , } 
\end{equation}
where the disk scaleheight $h(r)$ is related to the local sound speed $c(r)$ via
\begin{equation}
h(r) = \frac{c(r) }{\Omega_K(r) }\mbox{   . } 
\end{equation}
Integrating the density profile across the disk yields 
\begin{equation} 
\rho_m(r) = \frac{\Sigma(r)}{\sqrt{2\pi} h(r) }  \mbox{   . } 
\end{equation}
Meanwhile, the mid-plane gas pressure  is given by the equation of state (which we assume here to be a perfect gas), so 
\begin{equation}
p_m(r) = \frac{{\cal R} \rho_m(r) T(r) }{\mu } =  c^2(r) \rho_m(r) \mbox{   , } 
\end{equation}
where $\mu$ is the mean molecular weight of the gas and ${\cal R}$ is the gas constant.

In both cases, we need to know the local sound speed to evaluate $\rho_m$ and $p_m$. Following the standard $\alpha-$model for turbulent viscosity, we use
\begin{equation}
\nu(r)= \alpha c(r) h(r) \mbox{   , } 
\label{eq:alphamodel}
\end{equation}
where $\alpha$ is usually assumed to be constant in the entire disk.
Combining equations~(\ref{eq:nur}) and (\ref{eq:alphamodel}) we have 
\begin{eqnarray}
&& c^2(r) = c^2_{\rm AU} \left(\frac{r}{1{\rm AU}} \right)^{-1/2} \mbox{   where   } c^2_{\rm AU} = \frac{\nu_{\rm AU}}{\alpha} \Omega_{\rm AU} \mbox{   , } \nonumber \\
&& h(r) = h_{\rm AU} \left(\frac{r}{1{\rm AU}} \right)^{5/4} \mbox{   where   } h_{\rm AU} = \frac{c_{\rm AU}}{\Omega_{\rm AU}} \mbox{   . } 
\end{eqnarray}

Having laid out the governing equations for the disk, we need to choose its actual parameters. For the sake of comparison with the work of
\citet{Windmark2012}, we select the parameters for our disk to have the same properties at 1AU (see their Table 1). This immediately yields $\alpha = 10^{-4}$, $c_{\rm AU} = 10^5$cm/s, and the Reynolds number $\mathrm{Re} = 10^8$. Using a mean molecular weight of $\mu = 2.3$,
the temperature at 1AU is $276$K, close to the value reported by \citet{Windmark2012}. They choose a gas surface density of $\Sigma = 1700$g/cm$^{2}$,
which we can recover to a good approximation by selecting (for instance\footnote{Unfortunately, \citet{Windmark2012} do not report on the mass of the star used in their simulations; since the stellar mass determines the dynamical timescale of the disk, precise comparisons between our simulations and theirs is impossible.}), a central star of mass $M_\star = 0.75M_\odot$, surrounded by a disk of mass
$M_d = 0.05M_\star$, with a $R = 30$AU truncation radius. Using these parameters we find that the viscous timescale at 1AU is about 400,000 years, so
the gas disk would not evolve significantly during the simulation.

At 30AU, our disk model has a local gas surface density of 21g/cm$^{2}$ and a temperature of 50K. At this radius, the viscous evolution timescale is
larger than 10Myr, so again, the gas disk would not evolve significantly during the simulation. Finally, in all that follows we select a dust-to-gas
mass ratio of $Z = 0.01$. This ensures that the dust dynamics cannot influence the gas dynamics, as assumed throughout this work.

\subsection{Particle dynamics in the disk considered.}
\label{sec:particledynamics}

In what follows, we shall be interested in two representative disk regions, located close and far from the central star respectively. The close-in region is selected to be at 1AU, for ease of comparison with the work of \citet{Windmark2012}, and the far-out region is selected to be at 30AU, to address the problem of particle growth beyond cm-size discussed in Section~\ref{sec:intro}. 

\begin{figure}
\centerline{\includegraphics[width=0.5\textwidth]{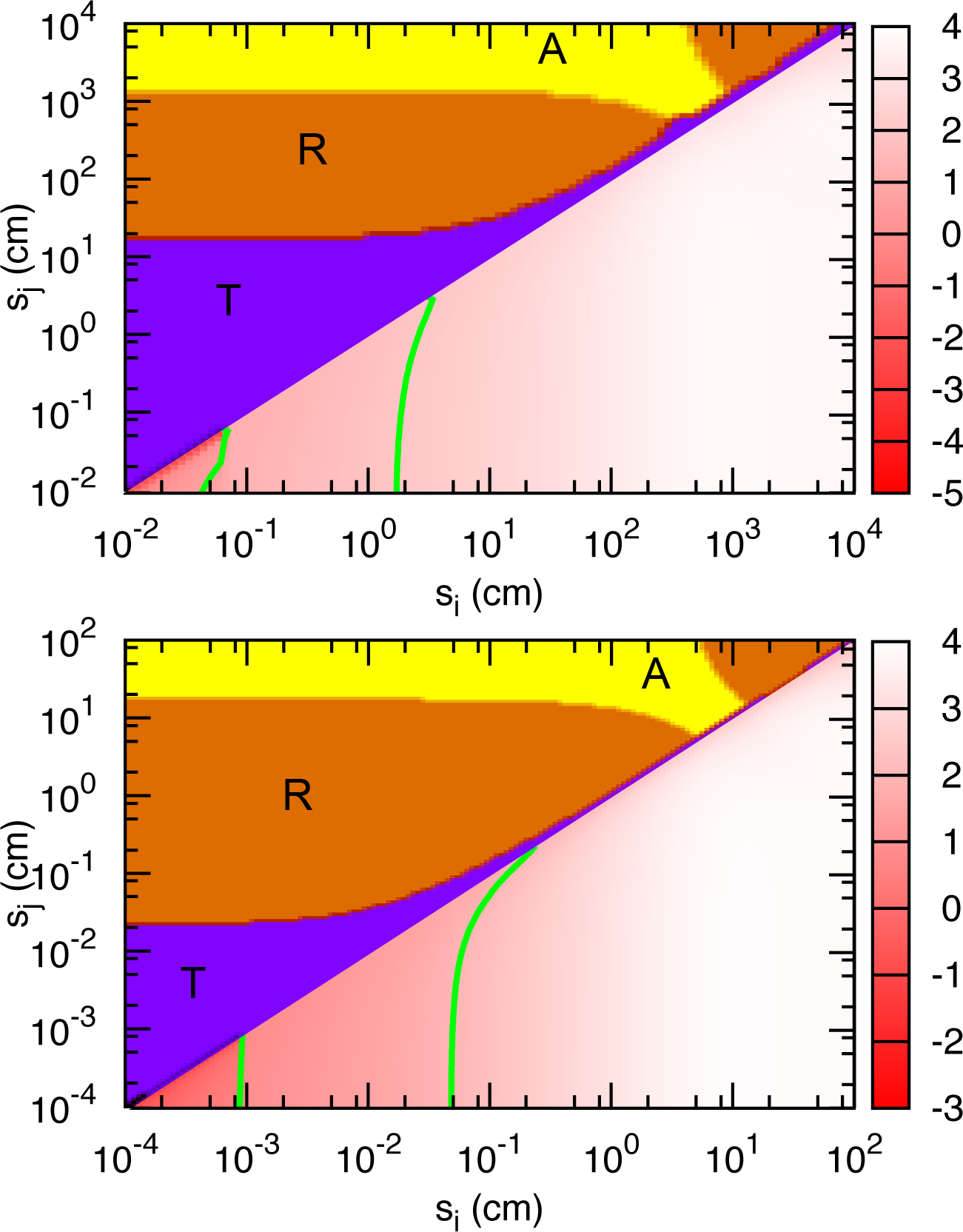}}
\caption{Collisional regimes and collisional velocities (top figure, 1AU, lower figure, 30AU). In each plot the top triangle shows the dominant
  collisional regime (purple = turbulence, orange = radial velocity, yellow = azimuthal velocity). Note that, for the disk regions and particle sizes
  considered, Brownian motion and vertical settling never appear to dominate. The lower triangle shows the logarithm of the actual mean relative
  velocity $\bar \Delta_{ij}$ as calculated in equation~(\ref{eq:dijevalue}), with the corresponding color bar shown on the right, as well as contours
  representing the bouncing and fragmentation thresholds: $\bar \Delta_{ij} = 5$cm/s (left) and $\bar \Delta_{ij} = 100$cm/s (right)
  respectively. }
\label{fig:regimesdv}
\end{figure}

As shown in Figure~\ref{fig:regimesdv}, relative particle motions are dominated by turbulence for sizes up to about 10 cm at 1AU. Collisions induced by differential radial drift dominate for larger particles. Hence, we expect our model to become more relevant than the one proposed by \citet{Galvagni2011} and \citet{Windmark2012} if growth occurs beyond 10cm at 1AU\footnote{Note that as radial drift becomes more important, the local model approximation unfortunately begins to fail; in this sense, the results of our model should only be considered indicative of processes which should be taken into account, rather than actual size predictions. See Section~\ref{sec:discuss} for detail.}. At 30AU, this transition happens for mm to cm size particles.

Figure~\ref{fig:regimesdv} also shows the logarithm of the mean relative velocity $\bar \Delta_{ij}$ of all $(i,j)$ particle pairs, and contours of
the bouncing and fragmentation thresholds $v_b = 5$cm/s and $v_f = 100$cm/s respectively. In what follows, we shall often refer to these contours as
the ``original'' bouncing and fragmentation barriers, as they often play that role in simulations which use the traditional method for kernel
construction (see Section~\ref{sec:oldkernels}). In any case, we find that bouncing is expected to affect particle growth beyond mm-size at 1AU, and
10 micron-size at 30AU, for the selected value of $v_b$. Fragmentation is expected to be important for particles around cm-size at 1AU, and mm-size
at 30AU.

\section{Results}
\label{sec:results}

We now proceed to study the collisional evolution of the particle distribution function, integrating equations~(\ref{eq:smolufrag}) numerically
at 1AU and 30AU in the disk, and compare the results obtained using three possible kernel construction procedures:
\begin{itemize}
\item Model O: The ``Old'' approach, in which kernels are constructed using equations~(\ref{eq:KernelB}), with $\bar \Delta_{ij}$ given by equation~(\ref{eq:tradbardelta}) and $\bar \epsilon_{ij}^s$ and $\bar \epsilon_{ij}^f$ given by equation~(\ref{eq:tradeps}).
\item Model M: The ``Maxwellian'' approach, in which kernels are constructed using equation~(\ref{eq:Maxwell}), with $\bar \Delta_{ij} $ given by equation~(\ref{eq:Maxwellmean}) and $\sigma_{ij}$ given by equation~(\ref{eq:sigmaij}).  This recovers in spirit (although not in detail) the idea proposed by \citet{Galvagni2011}, and \citet{Windmark2012}. 
\item Model N: The ``New''  approach, in which kernels are constructed using equations~(\ref{eq:fij}) and (\ref{eq:kij}). This treats mean motions and stochastic motions separately, as discussed in Section~\ref{sec:newkernels}. 
\end{itemize}
In all cases, the mean area cross-section is given by equation~(\ref{eq:aij}). 

In all simulations shown, the initial conditions are nearly mono-disperse, and are taken to be a narrow Gaussian centered on 
a mass-point that is slightly larger than the minimum mass-point
  considered. They have negligible influence on the results, except in the case of Model O when bouncing is included (see below for detail). 
Unless otherwise specified, the mass-mesh is logarithmically distributed and contains 6 points per decade in mass.  As discussed by \citet{Windmark2012} (see their Erratum), 
the resolution of the mass-mesh selected can affect the results significantly {\it on a quantitative} basis. A low resolution, as in any numerical method that uses finite differencing, introduces artificial dissipation 
which, in this particular case, manifests itself as diffusion in mass-space. The latter causes an artificial growth to larger sizes in low-resolution simulations. This is demonstrated in Appendix B, 
where we compare different runs at different resolutions in mass-space. Higher resolution runs have notably slower growth. However, they do reach qualitatively similar states as lower resolution runs, merely taking longer to do so. In this sense, a low resolution underestimates the timescale for growth to large sizes, but yields otherwise qualitatively correct answers. Since our goal here is qualitative rather than quantitative, we are satisfied with a fairly low resolution. We make sure that the latter is the same in all calculations, however, so that comparisons between different cases are meaningful. But our results {\it cannot} be taken to be quantitatively accurate, for this reason in particular and for all the other reasons discussed in this paper.

In what follows we plot the surface density distribution function $m \frac{d\Sigma}{dm} = h m^2 \frac{dn}{dm}$. As a result, if the mass distribution function evolves towards a 
power-law with $\frac{dn}{dm} \propto m^{-\lambda}$, then the surface density distribution function is proportional to $m^{2-\lambda}$.
At 1AU, as in \citet{Windmark2012}, we begin by ignoring the possibility of
bouncing, then include it. We finish by considering a model in which high-mass-ratio collisions lead to mass transfer from the projectile to the
target \citep{Teiser_Wurm_highVcoll,Kothe2010}, as in \citet{Windmark2012b} and \citet{Windmark2012}.  We then use the latter model to
  investigate the time evolution of the particle distribution function at 30AU as well.

\subsection{No bouncing, 1AU} 
\label{sec:nobounce}

\begin{figure}[h]
\centerline{\includegraphics[width=0.7\textwidth]{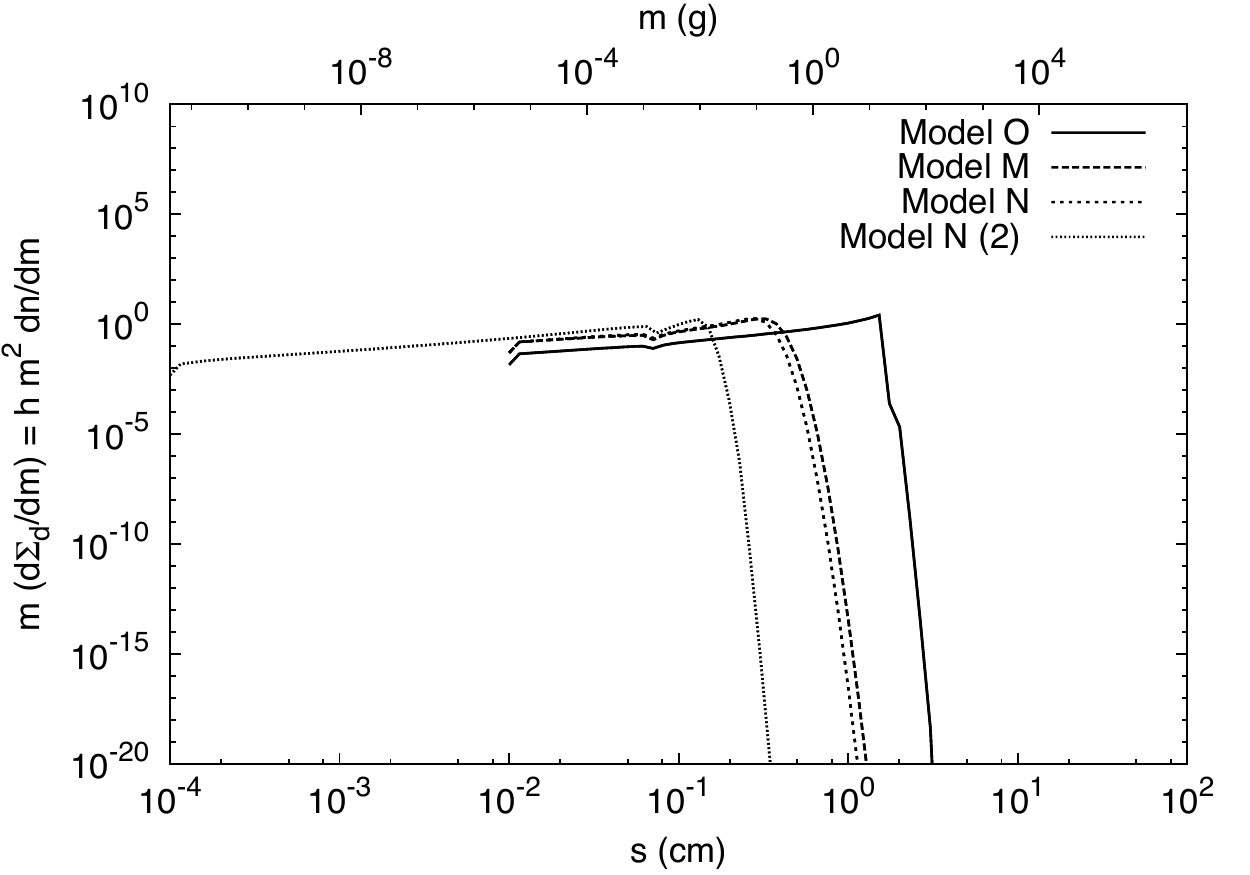}}
\caption{Particle evolution at 1AU, after 10,000 years, in models where the bouncing and fragmentation velocities are $v_b =v_f = 100$cm/s. Models O, N
  and M are described in the main text. In these simulations, the size-mesh spans the range $s \in [10^{-2}, 10^4]$cm (although the figure is
  truncated to focus on the region of interest). Model N (2) is the same as model N, but with a numerical mesh that extends down to a minimum particle
  size of $10^{-4}$cm instead of $10^{-2}$cm, with the upper mass point correspondingly reduced to $10^{2}$cm to keep the same resolution. }
\label{fig:vbvf-nomt}
\end{figure}

Let us first look at particle growth at 1AU, in the case where the possibility of bouncing is ignored (taking $v_b = v_f = 100$cm/s in all simulations). We compare the three possible evolution scenarios laid out above. We find that in all cases the particle mass distribution function rapidly evolves to a steady state (within less than 10,000 yrs), shown in Figure~\ref{fig:vbvf-nomt}. In each case, the steady state distribution $dn/dm$ is well-approximated by a truncated power law, with the same slope as that of the fragment mass distribution (see Section~\ref{sec:smolueqs}). The truncation size, however, depends on the model considered.

In Model O, particles grow up to 
a few centimeters in size. Predicting the maximum mass/size achievable in this model is in fact very easy -- one can simply read it at the intersection of the horizontal axis and the $\bar \Delta_{ij} = v_f$ contour\footnote{Technically, since this method only concerns Model O, the maximum size should be read from an equivalent plot showing the contour of $\bar \Delta_{ij} = v_f$ as calculated through equation~(\ref{eq:tradbardelta}) rather than (\ref{eq:dijevalue}); however, it happens that the two contours look very similar in the region considered.} in Figure~\ref{fig:regimesdv}. Indeed, since smaller particles are much more numerous than larger ones, high-mass-ratio collisions are much more common than equal-mass collisions. The growth of a large particle ``$i$'' is thus stalled when collisions with much smaller ones begin to lead to fragmentation instead of sticking. In Model O, where this transition occurs through a Heaviside function, 
the maximum particle size $s_i$ achievable is then simply given by $\bar \Delta_{ij} = v_f$, taking $j = 1$ to identify collisions with the smallest particles in the simulation.

As expected from the discussion in Section~\ref{sec:newkernels}, since particles are not seen to grow up to a size where radial drift becomes
important, the ``Maxwellian'' (M) model and the ``New'' (N) model predict very similar outcomes. In both cases, the maximum size achieved is
significantly smaller than in Model O. This is quite different from the results of \citet{Windmark2012}, and could be due to their possible use of 
``total'' probabilities instead of the mean fragmentation and sticking probabilities (see Section~\ref{sec:kernels}). Furthermore, we also find that the
maximum particle size achievable is sensitive to the minimum particle size considered in the simulation. This is illustrated in Figure
\ref{fig:vbvf-nomt} which shows, in addition to the three models described above, the outcome of a different run using Model N, with exactly the same
parameters and same resolution, but simply decreasing the minimum size considered by two orders of magnitude. The maximum particle size achieved in
this case is three times smaller than before.

After further investigation, we found that this effect is in fact generic to any model for which the fragmentation probability does not drop {\it exactly} to zero below a certain threshold velocity (which is the case in Models M and N, but not in Model O) -- and is in fact very easy to understand from a physical point of view. Indeed, as discussed above, collisions are much more frequent the smaller the projectile. In the very crude collisional model considered in this Section, it is possible (albeit unlikely) for a micron-size particle to collide with a much larger one and cause its fragmentation. Even though the fragmentation probability is very low, the sheer number of collisions that take place ensure that fragmentation does happen on a regular basis. Since the number density of particles is a steep function of their mass (or size), the new maximum particle size achievable thus sensitively depends on the smallest particle considered in the numerical integration, as well as the rate at which the fragmentation probability drops to zero when $\bar \Delta_{ij}$ drops below $v_f$.  This can be shown analytically, and will be the subject of a separate publication.

We note, however, that this trend disappears when considering models in which high-mass-ratio collisions do not lead to complete fragmentation, see Section~\ref{sec:masstrans} for details.  In this sense, our comments on the effect of the minimum particle size are somewhat academic.

 \subsection{With Bouncing, 1AU} 
\label{sec:bounce}

\begin{figure}[h]
\centerline{\includegraphics[width=0.7\textwidth]{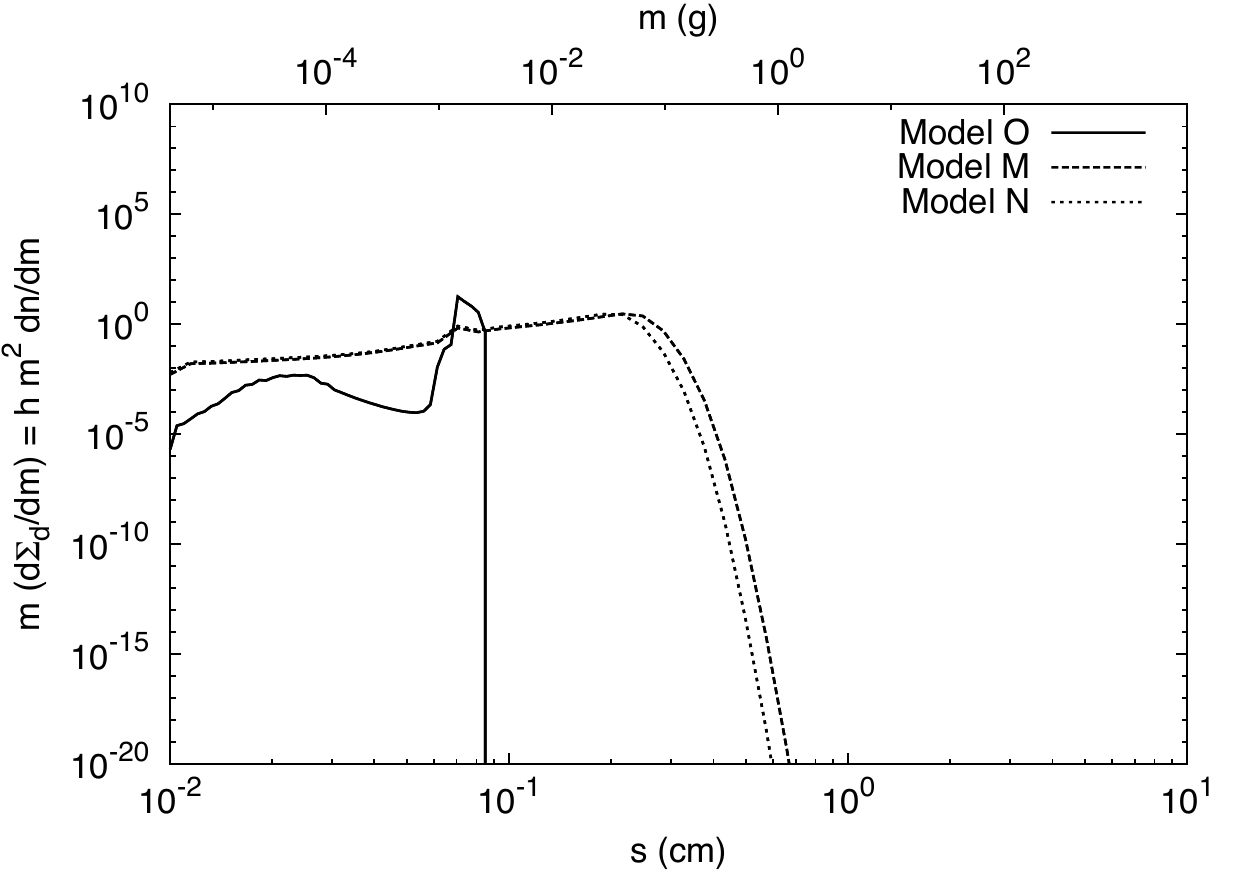}}
\caption{Particle evolution at 1AU, after 30,000 years, in models where the bouncing velocity is reduced to $v_b =5$cm/s compared with
  Figure~\ref{fig:vbvf-nomt}. The size range considered in Model O was reduced to $s \in [10^{-2}, 1]$cm, but with increased resolution (with 20
  mass-points per decade in mass) to capture the accumulation of particles near the bouncing barrier. Models M and N, by contrast have the same
  resolution and size range as in Figure~\ref{fig:vbvf-nomt} (with $s \in [10^{-2}, 10^4]$cm and 6 mass points per decade in mass), even though we
  only show sizes up to 10cm in this plot.}
\label{fig:vb5-nomt}
\end{figure}

We now consider the effect of bouncing, by setting $v_b = 5$cm/s. Again, we compare the three scenarios discussed above. We set the minimum particle size in the simulation to be $s_{\rm min} = 0.01$cm, for ease of comparison with \citet{Windmark2012} and with the results of the previous section, bearing in mind the potential effect of that choice on the outcome. In all cases, the simulation is integrated for 30,000 yrs, and the resulting particle size/mass distribution functions are shown in Figure~\ref{fig:vb5-nomt}.

At that point in time, Model O is still evolving. By construction, it will continue to do so until all particles have grown to a size above which any collisions they may undergo only result in bouncing. Little by little, all small particles disappear from the system. In many ways, this is the worst-case-scenario in terms of comparison with observations: small particles are entirely depleted, but growth to sizes larger than the bouncing barrier is also impossible. Note that the outcome in this case is strongly dependent on the initial conditions chosen. Suppose for instance that all particles are initialized {\it within} the bouncing region -- then, the particle size distribution function is in fact in a steady state. Here, we have selected to initialize all particles within the sticking region, to emphasize the effect of the bouncing barrier. 

By contrast with Model O, Models M and N have already reached a steady state by 30,000yrs. As in the previous section, the two are very similar to one
another, as expected from the fact that particles do not appear to grow beyond a size for which radial drift becomes important. As discussed by
\citet{Windmark2012}, the maximum particle size achieved is, this time, larger than in Model O since occasional low-velocity collisions are still
possible even when the mean relative velocity of the two particles is already beyond the bouncing barrier. More importantly, although fragmentation is
rare (since the mean fragmentation probability, at the sizes achieved, is very small), it is nevertheless sufficient to replenish the small-particle
population. In other words, models M and N provide more growth than Model O, and maintain a significant population of small grains, but the
  largest particle size achieved at steady-state is still much smaller than what is needed to form planetesimals. 

\subsection{With Bouncing and mass transfer, 1AU} 
\label{sec:masstrans}

As discussed in Section~\ref{sec:intro}, high-mass-ratio collisions are unlikely to result in the complete destruction of the larger body, as assumed
in Sections~\ref{sec:nobounce} and~\ref{sec:bounce}. Instead, the smaller particle (dubbed the ``projectile'' hereafter) is more likely to excavate a
small crater from the larger one (dubbed ``the target'' hereafter) and/or could itself partially or completely stick to the target \citep{Langkowski2008,Teiser_Wurm_highVcoll,Kothe2010} thus resulting in mass transfer.

Unfortunately, our understanding of the conditions under which mass transfer occurs, and in particular its dependence on the particle masses, porosity and velocities, is still in its infancy. For this reason, we use here a very simplified model of the process, in which we assume that the projectile completely sticks to the target if the mass ratio of two particles is larger than a certain critical value $\phi$ (which we take to be 50, by analogy with \citet{Windmark2012}), {\it and} if the collision would have otherwise led to fragmentation.
To implement this numerically, one simply needs to use the following algorithm: (1) Calculate the kernels $K_{ij}$ and $F_{ij}$ in the absence of ``mass transfer''. (2) If the mass ratio of the two particles is greater than $\phi$, first redefine $K_{ij}$ as $K_{ij} := K_{ij} +F_{ij}$, and then set $F_{ij} :=0$. 

This prescription differs from that of \citet{Windmark2012}, who assume that only 10\% of the mass of the projectile sticks to the target, and the
other 90\% fragments. We recognize that our own choice is likely to overestimate the actual amount of mass transfer occurring during a collision, but
since the actual value of $\phi$ is very poorly constrained anyway, the error made is within the same order of the approximation. We also 
show in Section \ref{sec:params} that the results presented below still hold qualitatively with lower mass-transfer rates. 
The advantage of this approach is that the mathematical structure of the effective
fragmentation/sticking probabilities is now very easy to visualize. For low mass ratio, they are as depicted in Figure~\ref{fig:epsij}. For high mass
ratio, $\bar \epsilon_{ij}^f = 0$ while $\bar \epsilon_{ij}^s$ now looks like the sum of the previously calculated sticking and fragmentation
probability functions (i.e. the sum of the two surfaces shown in Figure~\ref{fig:epsij}). In other words, $\bar \epsilon_{ij}^s$ is close to unity for
low collisional velocities {\it and} for high collisional velocities, with a substantial ``gap'' (or perhaps a ``moat'' may be more appropriate)
in-between that corresponds to the bouncing region. The latter is delimited by the original bouncing and fragmentation barriers (see Section
\ref{sec:particledynamics}). 

Mass transfer has fundamental implications for particle growth: as long as $\bar \epsilon_{ij}^s$ never strictly drops to zero anywhere, which is the
case in Models M and N as discussed in Section~\ref{sec:newkernels}, there is always a possibility for growth across the ``moat'' in high-mass-ratio
collisions. Indeed, while bouncing in this regime
may be the most likely outcome of a collision, there is always a possibility of very low- and very high-velocity events that result in coagulation and growth. 
Hence, once particles are large enough, they may always continue to grow by sweeping up smaller ones. As long as fragmentation via low-mass-ratio collisions is rare enough, the net effect is growth beyond the original barriers.

\begin{figure}[h]
\centerline{\includegraphics[width=0.7\textwidth]{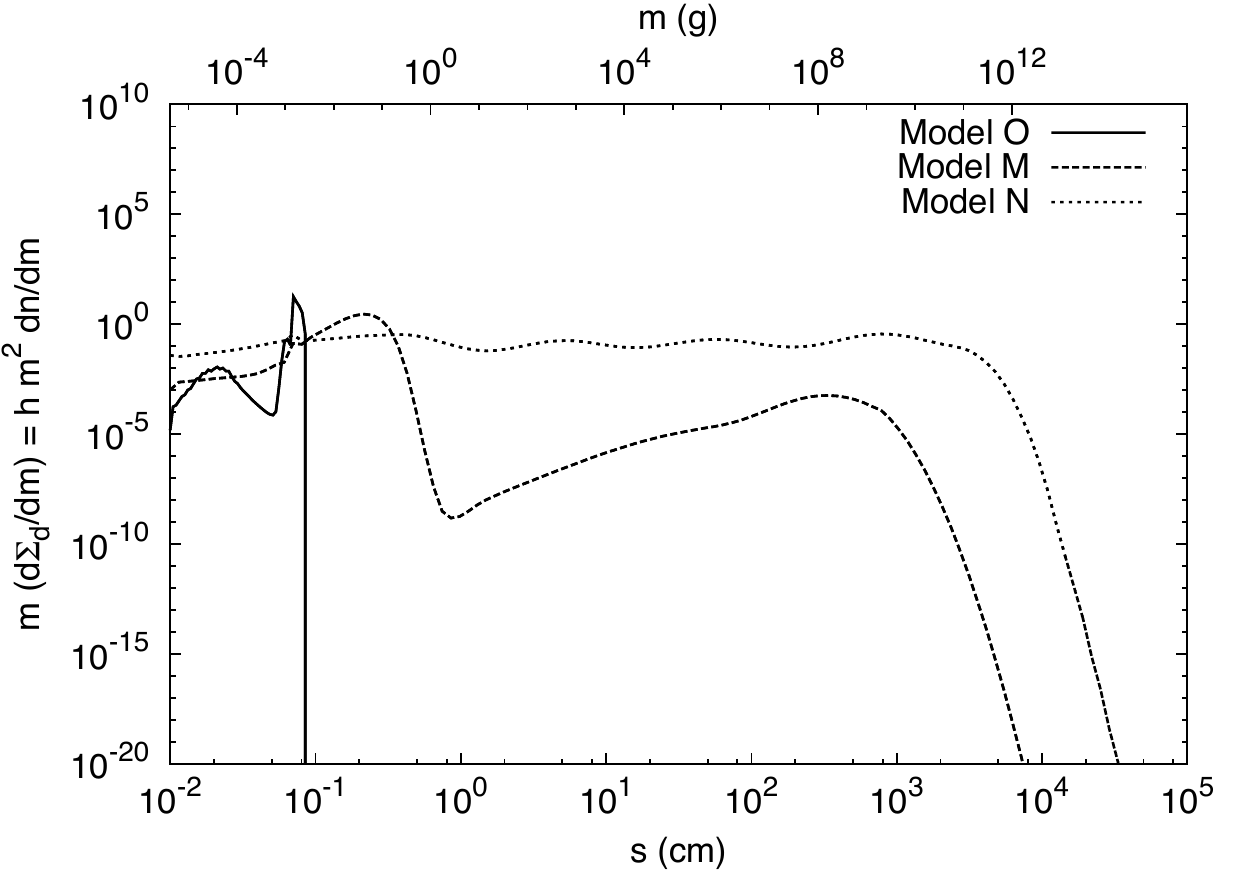}}
\caption{Particle evolution at 1AU after 10,000 yrs, with bouncing and mass transfer, in models where the bouncing velocity is $v_b =5$cm/s, and where
  the mass ratio above which every collision (that would otherwise lead to fragmentation) sticks is $\phi = 50$. The size range and resolution for
  Model O is the same as in Figure~\ref{fig:vb5-nomt}. The size range for Models M and N was increased to $s \in [10^{-2}, 10^7]$cm, with a
  corresponding increase in the number of mass-points to keep the same resolution.}
\label{fig:vb5-mt50}
\end{figure}

Figure~\ref{fig:vb5-mt50} presents the results of simulations that include the simple mass transfer model discussed above. We compare simulations using Models O, M and N, this time after 10,000 yrs only. The effect we are interested in is already very clear, and only becomes more pronounced beyond that time. One immediately notices that mass transfer has no effect on Model O. This was already pointed out by \citet{Windmark2012}, and is an obvious result in the light of the discussion above. Indeed, in Model O, both $\bar \epsilon_{ij}^s$ and $\bar \epsilon_{ij}^f$ strictly drop to zero when $\bar \Delta_{ij} \in [v_b,v_f]$. This means that, even for high mass-ratio collisions, $\bar \epsilon_{ij}^s$ strictly drops to zero for $\bar \Delta_{ij} > v_b$, preventing any possibility for growth beyond the original bouncing barrier. 

For Models M and N, however, we see a dramatic increase in particle growth. Furthermore, since
the largest particles are now in a regime that is dominated by radial drift rather than turbulence, the predicted evolution of the size distribution
functions for models M and N differ significantly. After only 10,000 years, the maximum particle size has increased by more than 4
orders of magnitude for Model M (corresponding to more than 12 orders of magnitude increase in mass, as found by
\citet{Windmark2012}). The effect is even more pronounced for Model N, where the maximum particle size is yet another order of magnitude
larger. Furthermore, the surface density of larger particles is 3-4 orders of magnitude larger in model N than in Model M. In short, correctly
modeling the difference between regular and stochastic processes when constructing the particle relative velocity distribution function turns out to
be highly beneficial to growth. Finally, note that even though fragmentation no longer completely suppresses the growth of large particles, it is
still amply sufficient to replenish the small grain population.

\subsection{Evolution of the particle distribution function at 30AU} 
\label{sec:masstrans30AU}

Using Model N, we can now study other regions of the disk. Here, we consider the disk at 30AU, and include both bouncing and mass transfer with the same parameters as in the previous sections: $v_b = 5$cm/s, $v_f = 100$cm/s, and $\phi =50$. As seen in Figure~\ref{fig:regimesdv}, the bouncing threshold now occurs for much smaller particles than at 1AU. For this reason, we shift the mass-mesh so that $s \in [10^{-4},10^5]$cm instead, and keep the same resolution. 

\begin{figure}
\centerline{\includegraphics[width=0.7\textwidth]{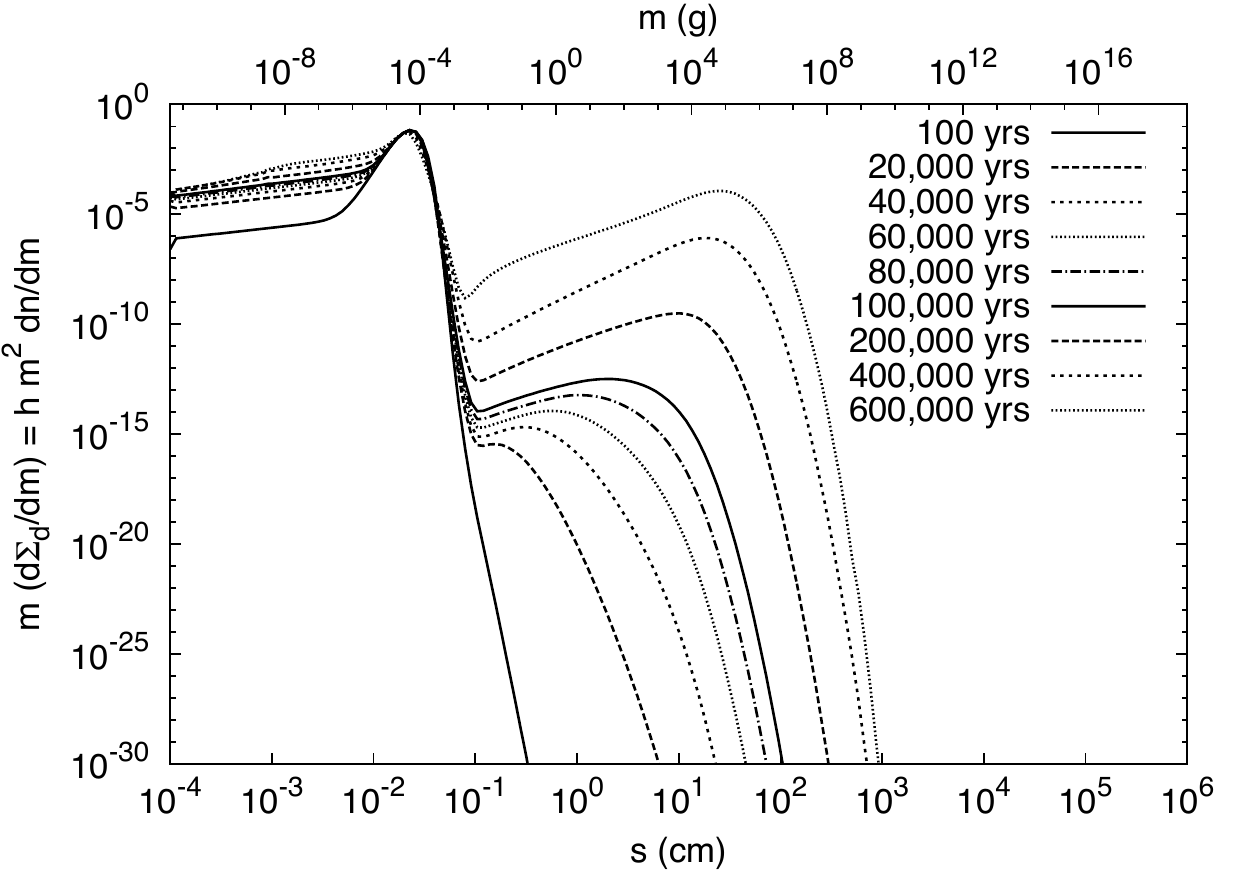}}
\caption{Particle evolution at 30AU with bouncing and mass transfer using model~N, where the bouncing velocity is $v_b =5$cm/s,
  and where the mass ratio above which every collision (that would otherwise lead to fragmentation) sticks is $\phi = 50$. This figure shows snapshot
  of the mass distribution function at different times. Note the gradual emergence of a large particle population, of sizes up to a few cm. Although difficult to see, mass is indeed conserved in this simulation -- as time evolves, the particle peak is slightly eroded. }
\label{fig:evolve-30AU}
\end{figure}

In order to present complementary yet comparable information to that of the previous sections, we show the actual temporal evolution of the
particle distribution function. It is clear from Figure~\ref{fig:evolve-30AU} that the system now contains two interacting but distinct particle
populations. The small particle population is distributed as a truncated power law, with a power index dictated by the assumed fragmentation law of
the model (here, with $dn/dm \sim m^{-1.83}$, see Section~\ref{sec:smolueqs}), and a maximum size around 0.1 mm. At early times, the bouncing barrier
induces a pile-up just above that size, giving the appearance of a fairly mono-disperse population. However, the pile-up gradually disappears later
on, as the fragmentation of an increasing number of larger particles replenishes the small particle population. Meanwhile, the larger particles 
steadily sweep up the smaller ones, resulting in the gradual emergence of a large-particle population. 

After about 500,000 years (in this particular simulation), the large particle population has become very significant indeed.  Its distribution, especially at later times, can also be
viewed as a truncated power law. The power index is shallower than that of the small particles, with $dn/dm \sim m^{-1.2}$. The maximum particle size
within the large-particle population initially grows quickly with time, increasing by 3 orders of magnitude in the first 100,000 years. Later on,
the growth of the maximum particle size slows down in favor of a steady increase in the {\it number density} of the
large particles while keeping the same overall shape of the distribution.

It is clear from this simulation that the effects described here have the potential of solving simultaneously the two important puzzles raised
by observations of protoplanetary disks -- the persistence of a small-particle population for millions of years, and the
inferred presence of rather large particles ($>$cm size) far out in the disk. Furthermore, the emergence of two particle
populations with different mass distribution functions, rather than a single continuous power law, is consistent with the observations of
\citet{Wilner2005}. We will return to these points in Section~\ref{sec:discussobs}. At this point, however, it is time to look into the dependence of
the model results on the input parameters.

\subsection{Dependence on parameters} 
\label{sec:params}

\begin{figure}
\centerline{\includegraphics[width=0.7\textwidth]{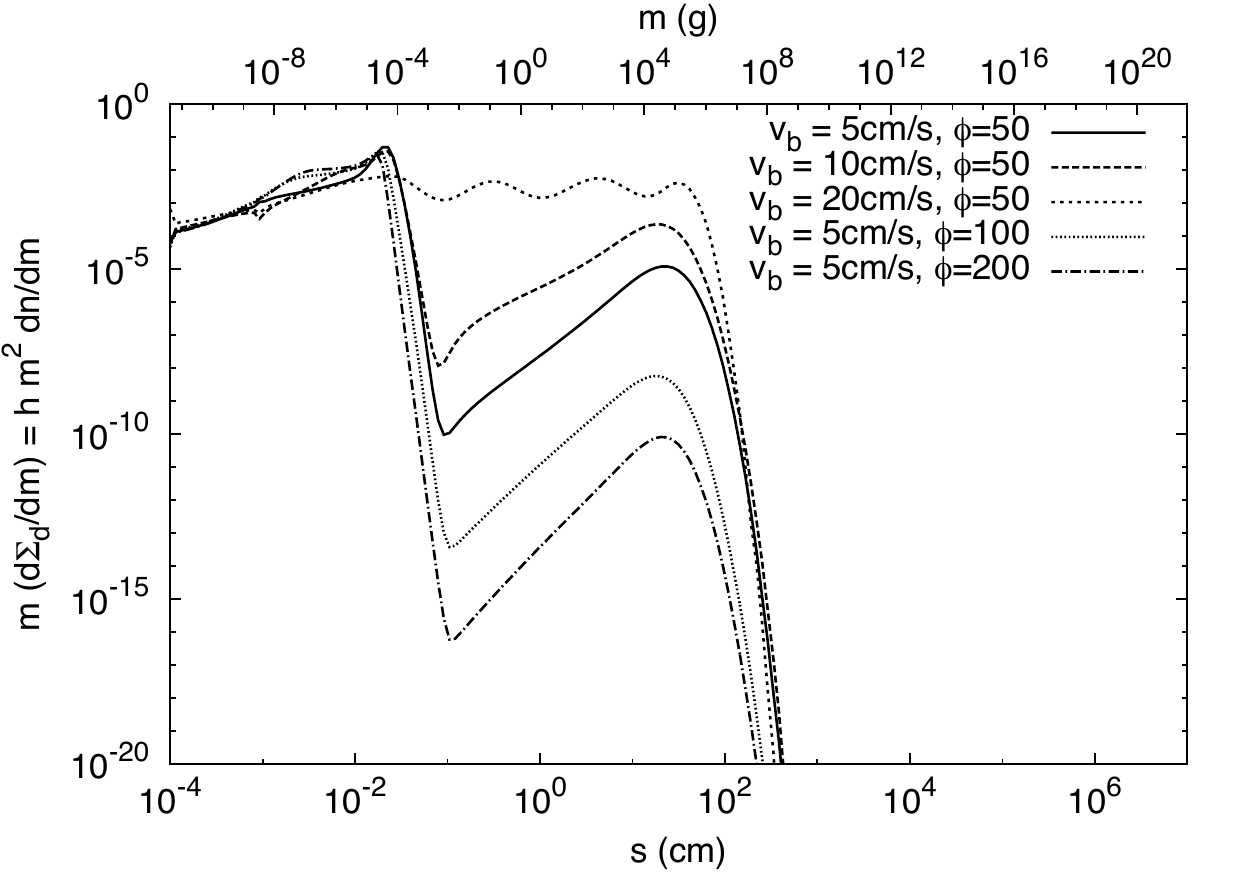}}
\caption{Particle evolution at 30AU with bouncing and mass transfer, in models with varying bouncing velocity $v_b$ and mass ratio above which every
  collision sticks, $\phi$. This figure shows snapshot of the mass distribution function at 500,000yrs. Note that all models are still evolving at the given time. }
\label{fig:model-compare}
\end{figure}

In all previous simulations, we used the same values for the fragmentation threshold $v_f$, the bouncing threshold $v_b$, and the mass ratio above which mass transfer happens. We now vary the latter two to see their effects on evolution of the particle distribution function. The results are presented in Figure~\ref{fig:model-compare}.

We consider here the same disk at 30 AU (very similar results apply at 1AU). In all simulations, we fix $v_f = 100$cm/s, but vary $v_b$ between 5cm/s
and 20cm/s, and $\phi$ between 50 and 200. We evolve the system of equations~(\ref{eq:smolufrag}) for 500,000 years. In all cases we observe the same
qualitative evolution of the particle size distribution function into the two particle populations described in the previous section (small particles
and large particles). However, we also see that the transfer of mass from the small to the large particle population, which in turn controls the large
particle population growth rate, depends quite sensitively on $\phi$, and on the width of the bouncing region (i.e. on $v_f - v_b$).  

First, we find that the larger $\phi$ is, the slower the mass flux from small to large particles. This effect is intuitive: a given target mass
$m_i$ will effectively sweep all particles of mass $m_i/\phi$ or smaller. The total amount of mass available for growth can be evaluated from
$\int_0^{m_i/\phi} m (dn/dm) dm$, and is naturally smaller when $\phi$ increases (the details of exactly how much smaller depends on the specific
shape of the mass distribution function for small particles).

\begin{figure}
\centerline{\includegraphics[width=0.7\textwidth]{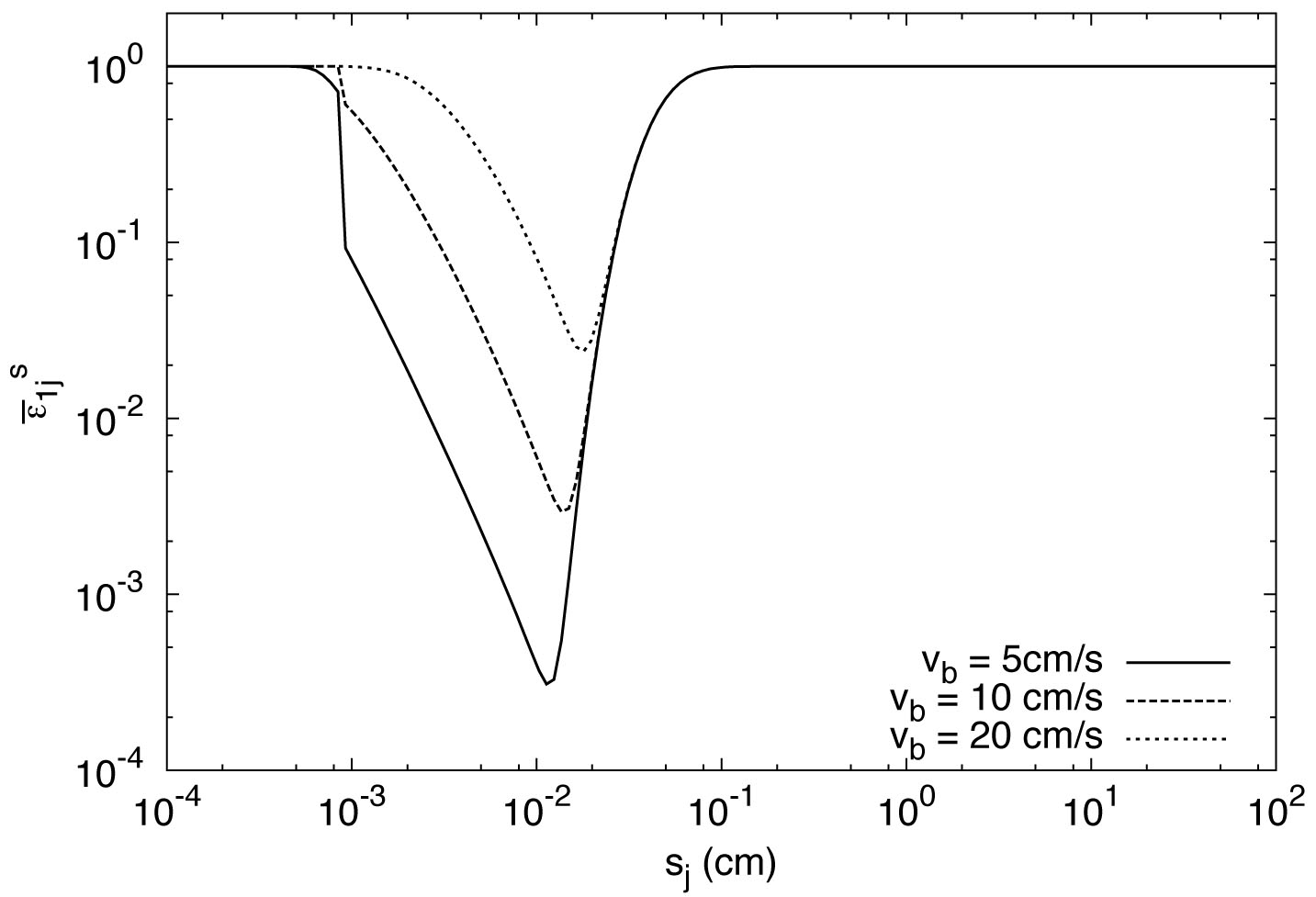}}
\caption{Effective sticking probability $\bar \epsilon_{1j}^s$ of a particle of index $i=1$ (the smallest possible particle) with a particle of
  size $s_j$, for $\phi = 50$, $v_f = 100$cm/s and various values of $v_b$ at 30AU. The wider the bouncing region, the deeper the minimum in the
  function $\bar \epsilon_{1j}^s$. The apparent discontinuity for $s_j \simeq 10^{-3}$cm is due to the non-smooth nature of the \citet{OrmelCuzzi2007} turbulent velocity prescription. }

\label{fig:gap}
\end{figure}

This mass flux is also dependent on the width of the bouncing gap.
To understand this, let's consider for instance the tiniest particles in the system (taking $i = 1$), and look at their sticking probability with
other particles around. In a mass transfer scenario, collisions with anything of mass $m_j > \phi m_1$ do not lead to fragmentation. For $\phi = 50$
and $s_1 = 10^{-4}$cm, this corresponds to $s_j > 4 \times 10^{-4}$cm -- in other words, very few of the collisions involving $s_1$ lead to
fragmentation. However, the mass transfer model considered here only leads to sticking for collisions that would otherwise lead to
fragmentation. Bouncing remains a barrier to growth, and thus the larger the bouncing region, the stronger this barrier is. Figure~\ref{fig:gap}
illustrates this clearly, by showing the mean sticking probability $\bar \epsilon_{1j}^s$ of a particle of index $i=1$ colliding with a particle of size
$s_j$. The wider the gap, the deeper the minimum in the function $\bar \epsilon_{1j}^s$. This minimum is the growth bottleneck of the system, and thus
controls the flux of mass from the small particle population to the large particle population.

\section{Discussion and conclusions}
\label{sec:discuss}

In the previous sections, we presented a new model for the construction of the coagulation and fragmentation kernels involved in the evolution of the
particle size distribution function, which takes into account the p.d.f.s of the relative velocities of the particles. In particular, we include
  a more physically motivated approach which separates the deterministic from the stochastic relative particle velocities. When combined with the possibility that high-mass-ratio collisions can lead to the growth of the (high-mass) target rather than its partial or complete destruction, 
this model is potentially able to solve three of the longest-standing problems raised by observations of protoplanetary disks: the persistence of small-grains for millions of years despite ongoing coagulation, the existence of cm-size particles far out in the disk, and finally, the formation of large planetesimals close to the central star inferred from the ubiquity of extrasolar planets. 

In this section, we now discuss the model in more detail, and in particular, which of these results are model-dependent, and which are not. We then review our result in the light of observations. 

\subsection{Model dependence}

The results of Section~\ref{sec:results} suggest that growth beyond the traditional ``bouncing'' and ``fragmentation'' barriers, {\it and} the
maintenance of a small-particle population, merely requires two ingredients: (1) a threshold above which high-mass-ratio collisions lead to sticking
rather than to fragmentation, and (2) mean sticking and fragmentation probabilities that never strictly drop to zero.

In fact, these ingredients are always naturally expected to arise in any ``physically motivated'' model for the evolution of the particle size distribution function. Their absence from previous work can only be viewed as unfortunate and misleading oversimplifications of the problem. 

Let us begin by considering the case of high-mass-ratio collisions. In the original work of \citet{DullemondDominik2005} for instance, the fragmentation and sticking probabilities were constructed by considering the total kinetic energy of the collision in comparison with the binding energy of the particle. In these kinds of models, high-mass-ratio collisions are much less likely to lead to the fragmentation of the larger body, and sticking becomes a more plausible outcome. In some sense, such prescriptions already incorporate the idea of mass transfer, without any need to define it artificially. Unfortunately, much of the work published between 2005 and 2011 dropped this energy-dependent view of the fragmentation/sticking thresholds in favor of a velocity-dependent one (such as the one used, for the sake of illustration, in this paper). We advocate that future work should return to using the energy-based approach in which the notion of mass transfer naturally emerges, in conjunction with the use of velocity p.d.f.s for the construction of the kernels.

Mass transfer on its own, however, is not sufficient: the second condition is equally important. This was illustrated in Section~\ref{sec:masstrans}, where we showed that even with mass transfer Model O does {\it not} lead to growth beyond the fragmentation barrier. In other words, in any model where there exists a given particle size $s_i$ for which the mean sticking efficiency $\bar \epsilon_{ij}^s =0$ for {\it all} possible particle pairs  $(i,j)$, growth beyond $s_i$ is naturally impossible. This was a rather common outcome of models that used the ``traditional approach'' in conjunction with piecewise defined functions $\bar \epsilon_{ij}^s$ and $\bar \epsilon_{ij}^f$ that drop strictly to zero across a particular velocity threshold, as in the work of \citet{Brauer2008} and the piecewise linear model used by \citet{Birnstiel2010}. 

In reality, however, the mean sticking probability for a given particle size is never expected to drop to zero entirely. 
Even when the mean collisional velocity/energy of a particle pair $(i,j)$ is high, low velocity/energy collisions
are always possible. This is explicitly taken into account when using p.d.f.s of relative velocities in the calculation of the mean sticking and fragmentation probabilities (see Section~\ref{sec:newkernels}). Indeed, since the latter can be written as convolution products of the relative velocity p.d.f. and the sticking and fragmentation probabilities of individual collisions, they are always strictly positive. Again, we 
advocate that such a model should always be used in the future. 

As long as it is using the two ingredients listed above, we expect that any {\it local} model 
will yield answers that are similar to the ones shown in 
Section~\ref{sec:results} for Model N, and will reveal the emergence of two populations of particles: a small-particle population constantly replenished by the fragmentation of larger bodies (with a size distribution function controlled by a collisional fragmentation cascade), and a large-particle population that grows by sweeping the smaller particles (with a size distribution controlled by a coagulation/fragmentation balance). In this sense, anyone's preferred prescription for the relative velocity p.d.f.s, for the mass transfer scenario, for the effect of porosity on the collisions, should yield qualitatively similar results -- the latter are certainly not model-dependent. 

However, we also showed that the details of the mass flux between the two populations are, unfortunately, very much dependent on the model considered, and within a given choice of model, on the selected parameters. We illustrated this in Section~\ref{sec:params} by considering a range of parameters within our own toy model, and showed that the predicted surface density of large particles in the system can vary by 5-10 orders of magnitude simply by changing the parameters by a few. This somewhat unsatisfactory result can be understood, as shown earlier, by the fact that the mass flux between the small and large particle populations is ultimately controlled by the sticking efficiency bottleneck, as  well as the available surface density of small particles that a larger one can ``sweep'', which are themselves strongly dependent on the parameters.

Furthermore, we showed in Appendix B that the results are, unfortunately, also sensitively dependent on the numerical resolution used to discretize the mass-mesh, if the latter is too small. In particular, a low resolution appears to shorten the growth timescale artificially, by causing an artificial diffusion in mass-space of the particle size distribution function. Interestingly, however, we find that higher resolution runs eventually do reach very similar states as the lower resolution ones, but take longer to do so. In this sense, the emergence of two particle populations is robust. We also find that the typical particle sizes achievable in both populations is very similar in high- and low-resolution runs, demonstrating that this is again a generic property of the system.

As discussed in Section~\ref{sec:intro}, the model presented in this work was, by and large, selected for simplicity rather than physical realism. Comparing results directly with observations will require much more sophistication in the physics included. The particle structures (porosities, composition) should be taken into account. The sticking and fragmentation probabilities for a given collision are 
more likely to depend, in a complex manner, 
on the collisional energy and the material strength of the two particles. This should also be taken into account instead of using equation
(\ref{eq:epsbasic}). In addition, the relative velocity p.d.f.s of two particles undergoing turbulent motion is not likely to be a Maxwellian, as we
had to assume here. Indeed, the velocity p.d.f. of a single particle interacting with turbulent eddies is known to be better represented by Levy-type
distributions than by Gaussian/Maxwellian distributions. The collisional relative velocities p.d.f.s could therefore differ significantly from the one given in equation (\ref{eq:newpdf}). Levy-type distributions are defined by much more slowly decaying tails; this is likely to affect the
particle size distribution evolution significantly (although quantitatively rather than qualitatively).

In addition, we must remember that the models studied in Section~\ref{sec:results} are local models, in the sense that they neglect the radial flux of particles in and out of the domain considered. This approximation is adequate a long as the mass flux in and out of a mass bin caused by radial drift is small compared with that caused by coagulation and/or fragmentation. As we saw, including mass transfer and velocity p.d.fs is always sufficient, in a local model, to trigger the growth of a large-particle population. In a global disk model, however, this effect will have to be sufficiently strong to overcome the effect of radial drift on the large particle growth. Interestingly, however, we now understand what controls the growth rate beyond the fragmentation barrier, so the observations 
of the ubiquity of exoplanets, as well as of the presence of a cm-size particle population at large radii, are still reconcilable with our type of model, and will help place constraints on assumed sticking and fragmentation parametrization and on the particle velocity p.d.f. selected. 

Finally, note that taking into account the effect of radial drift in our velocity p.d.f.s enhanced particle growth significantly compared with the similarly local Maxwellian-only type of model proposed by \citet{Windmark2012}, in particular in the cases with mass transfer. This is because, for high-enough mass ratio between the two particles, high velocity collisions (which are more frequent when radial drift is taken into account) result in sticking. As such, it is much more likely to yield growth to large sizes despite drift than theirs, and should therefore be used preferentially.

\subsection{Observational perspective}

\label{sec:discussobs}

  Various observational studies of protoplanetary disks have inferred the presence of a wide distribution of particle sizes from submicron-sized
  dust grains to cm-sized pebbles
  \citep{1989ApJ...340L..69W,1996A&A...309..493D,1998Natur.392..788H,1999osps.conf..579B,2000AJ....120.3162L,Wilner2005,2006ApJ...638..897S,2008Ap&SS.313..119A,Lommen2009}. Intriguingly,
  no conclusive evidence has been found for a correlation between the age of primordial disks and the properties of their population of small dust
  grains \citep[as revealed by silicate features:][]{2006ApJ...639..275K,2009ApJ...703.1964F,2010ApJ...714..778O,2010A&A...521A..66R}. \citet{2011ARA&A..49...67W}
  attribute this to a balance between grain growth and destruction on the one hand, and between crystallization and amorphization on the other, concluding that this balance seems
  to persist throughout the duration of the primordial disk stage (at least in the surface layers of the disk). The consequence for the scenario in
  which planet formation proceeds via dust growth is the coexistence of particles growing to planetary sizes with a population of small grains that
  are resupplied by continuous fragmentation. Future observations of planets embedded in a young protoplanetary disk or the confirmation of candidate
  planets in transition disks, such as T Cha \citep{2011A&A...528L...7H} or LkCa15 \citep{2012ApJ...745....5K}, would support this picture.

  In the meantime, other proxies have been sought for evidence of ongoing planetary formation.  Unfortunately, the failure of theoretical collisional models to
  produce results in which large and small particles co-exist have, until now, hindered these efforts. The new model presented in this paper
  paves the way to resolving this problem.  Our physically motivated coagulation and fragmentation kernels are able to simultaneously establish a sustained growth of
  particles and to preserve a micron-sized dust population in a protoplanetary disk. The particle
  size distribution that naturally emerges is one in which the small and large particle populations have notably different power-laws: one
  that is dominated by fragmentation, and one that is dominated by coagulation via sweeping. Since the upper size-cutoffs and total mass in each population are strongly dependent on the model considered, observations may help to rule out certain aspects of the parameter space that affect the theoretical results, enabling the model to be constrained.

  From a qualitative point of view, the existence of two particle populations far from the central star agrees well with detailed modeling and
  cm-observations of the TW Hydrae disk by \citet{2002ApJ...568.1008C} and \citet{Wilner2005}. 
  Furthermore, we expect large and small particles to coexist at different radial positions in the disk, albeit with a strong variation in the maximum size achievable by the larger particles -- 
  in other words, one may anticipate strong radial gradients in inferred particle growth beyond mm-size. This result is interesting in the light of the work of \citet{2011A&A...529A.105G}, who
  reported observational evidence for a radial dependence of the grain size in protoplanetary disks. We expect that more detailed future observations
  probing different parts of a disk will be able to determine whether or not two particle populations do indeed coexist at multiple radii in a disk,
  as suggested by our model. In any case, in light of our results, observers should always consider the possibility that two populations of
  particles exist when fitting spectral energy distributions of disks, rather than a single continuous population from small to large sizes.

\acknowledgements This project was initiated during the ISIMA 2011 summer program hosted at the Kavli Institute for Astronomy \& Astrophysics in
Beijing, and was funded by the NSF CAREER grant 0847477, the NSF of China, the Silk Road Project, the Excellence Cluster Universe in Munich,
Theoretical Astrophysics at Santa Cruz, and the Center for Origins, Dynamics and evolution of Planets. We thank them for their support.
P.G. acknowledges support by NSF CAREER grant 0847477.  FM also acknowledges
  the support of the German Research Foundation (DFG) through grant KL 650/8-2. MG acknowledges support from the University of Zurich PhD
program. C.O. appreciates funding by the German Research Foundation (DFG) grant OL 350/1-1. FM is
  supported by the ETH Zurich Postdoctoral Fellowship Program as well as by the Marie Curie Actions for People COFUND program.

\bibliographystyle{apj}
\bibliography{Disks}

\appendix

\section{Appendix A: Derivation of the relative velocity distribution function }

To derive equation~(\ref{eq:newpdf}) from (\ref{eq:p3d}), we expand ${\mathbf \Delta}_{ij}$ in spherical coordinates (to be specified, see below) as
\begin{eqnarray}
\Delta_{r,ij} &=& \Delta_{ij} \sin \theta \cos \phi \mbox{   , } \nonumber \\
\Delta_{\varphi,ij} &=& \Delta_{ij} \sin \theta \sin \phi \mbox{   , } \nonumber \\
\Delta_{z,ij} &=&\Delta_{ij} \cos \theta \mbox{   , } 
\label{eq:spherexp}
\end{eqnarray}
and integrate equation~(\ref{eq:p3d}) over a spherical surface of radius $\Delta_{ij}$: 
\begin{equation} 
p(\Delta_{ij} ) d \Delta_{ij}  = \int_0^\pi \int_0^{2\pi}p_{3D} ({\mathbf \Delta}_{ij} ) \Delta_{ij}^2 \sin\theta d\theta d\phi \mbox{   . } 
\end{equation}
Assuming isotropy in the particle dispersion (see Section~\ref{sec:newkernels}), 
one can re-write $p_{3D} ({\mathbf \Delta}_{ij} ) $ as 
\begin{equation} 
p_{3D}({\mathbf \Delta}_{ij}  ) = \frac{1}{(2\pi)^{3/2} \sigma_{ij}^3 } \exp\left( - \frac{\Delta_{ij}^2 + (\bar \Delta^D_{ij})^2 }{2\sigma_{ij}^2} + \frac{{\mathbf \Delta}_{ij} \cdot \bar {\mathbf \Delta}^D_{ij}  }{\sigma_{ij}^2} \right) \mbox{   , } 
\end{equation}
where $\bar {\mathbf \Delta}^D_{ij} = ( \bar u_i - \bar u_j, \bar v_i - \bar v_j, \bar w_i - \bar w_j) $ and $\bar \Delta^D_{ij} = |\bar  {\mathbf \Delta}^D_{ij}|$ (see equation~(\ref{eq:meanvel})). The integral over the spherical surface of $p_{3D}({\mathbf \Delta}_{ij})$, written in this form, is much simpler if the expansion in spherical coordinates (\ref{eq:spherexp}) is cleverly chosen with $\bar {\mathbf \Delta}^D_{ij}$ defining the polar axis, so that ${\mathbf \Delta}_{ij} \cdot \bar {\mathbf \Delta}^D_{ij}  = \Delta_{ij} \bar \Delta^D_{ij} \cos\theta$. We then have
\begin{equation}
p(\Delta_{ij} ) d \Delta_{ij}  = \frac{2\pi \Delta_{ij}^2 }{(2\pi)^{3/2} \sigma_{ij}^3 } \exp\left( - \frac{\Delta_{ij}^2 + (\bar \Delta^D_{ij})^2}{2\sigma_{ij}^2} \right) \int_0^\pi \exp\left( \frac{\Delta_{ij} \bar \Delta^D_{ij} \cos\theta}{\sigma_{ij}^2} \right) \sin\theta d\theta \mbox{   , } 
\end{equation}
which can be integrated to recover (\ref{eq:newpdf}).

\section{Appendix B: Effect of resolution on the simulation results}

We compare here the outcome of four simulations run using Model N (see Section \ref{sec:masstrans})
but with different mass-mesh resolutions. To avoid impossibly long integration times, we have selected a model that does lead to rapid growth in general, by using a low value of the mass transfer ratio ($\phi = 30$), and a fairly narrow gap (with $v_b = 20$cm/s, $v_f = 100$cm/s), and have selected our radial location to be 1AU. The four simulations compared have 150, 300, 600 and 900 mass-points respectively, corresponding to 6, 12, 24 and 36 mass bins per decade. At a given time, here $t=1,500$yrs, the higher-resolution runs have experienced significantly less growth than the lower-resolution run, but their particle size distribution functions eventually become qualitatively similar (i.e. similar ``large-particle'' peak position and height) to that of the lower-resolution run a little later in time. In this sense, the results from this paper (and all others using similar methods) can only be taken to be qualitatively correct, and should not be interpreted as strict predictors of the particle sizes and growth timescales in accretion disks. However, since our main argument here compares different models using the same resolution, the comparison is meaningful.

\begin{figure}
\centerline{\includegraphics[width=0.7\textwidth]{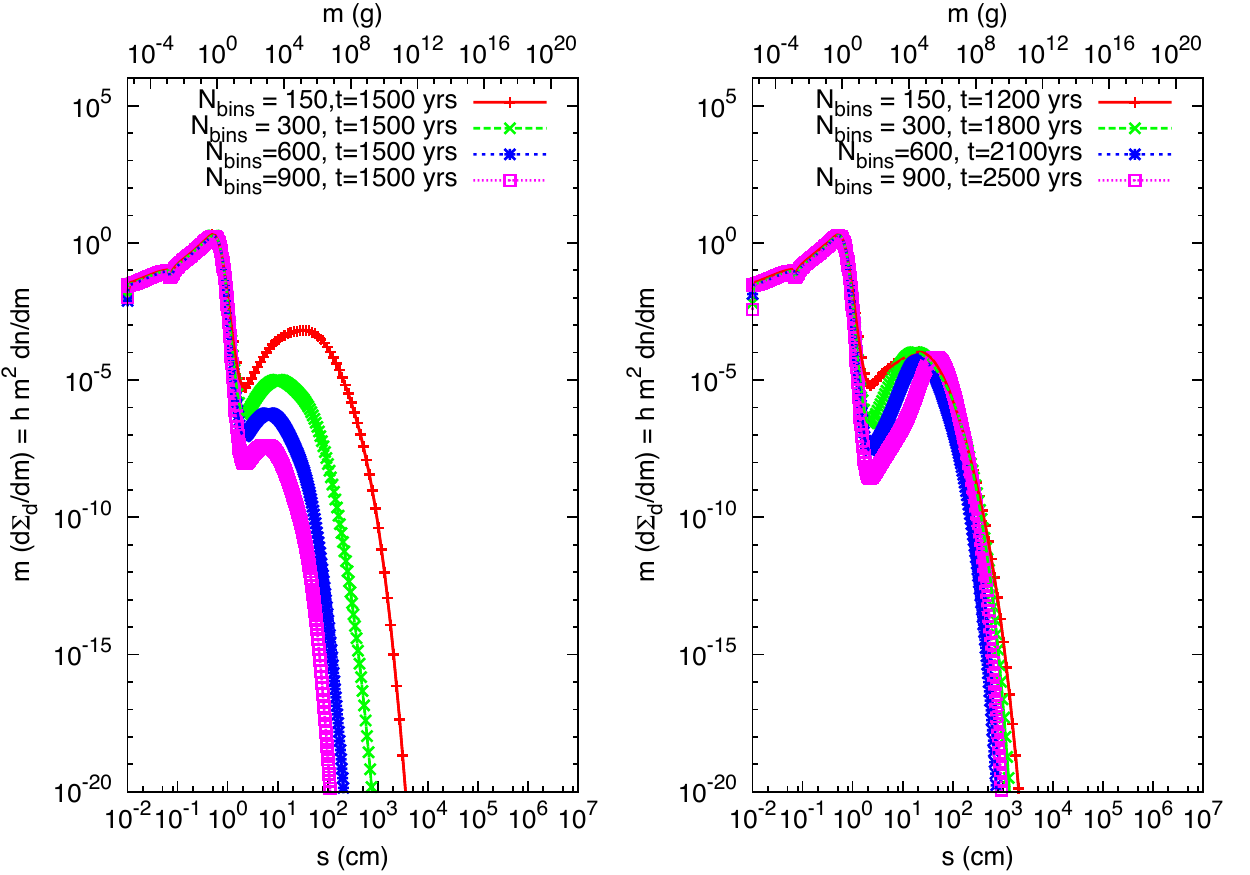}}
\caption{Model N with $v_b = 20$cm/s and $\phi = 30$, with minimum particle size $s_{\rm min} = 10^{-2}$cm and maximum particle size $s_{\rm max} = 10^7$cm. Left: Evolution of the system after 1,500 yrs, for 4 different resolutions, as measured by the total number of mass bins considered. The lowest resolution is the one used in this paper, with about 6 mass bins per decade. Higher resolutions (2, 4 and 6 times higher) are shown for comparison. At a given time, particle growth beyond the bouncing barrier is qualitatively similar, but clearly significantly reduced in the higher resolution runs. Right: Higher resolution models do lead to similar particle growth, but on a longer timescale.}
\label{fig:resolution}
\end{figure}

\end{document}